\documentclass[10pt,conference]{IEEEtran}
\IEEEoverridecommandlockouts

\usepackage{amsmath,amssymb,amsfonts,amsthm}
\usepackage{bbm}
\usepackage{dsfont}
\usepackage{cite}
\usepackage{graphicx}
\graphicspath{ {./figures/} }
\usepackage[linesnumbered,ruled,vlined]{algorithm2e}

\makeatletter
\def\input@path{{./figures/}}
\makeatother

\theoremstyle{plain}

\DeclareMathOperator*{\E}{\mathds{E}}
\DeclareMathOperator*{\prob}{\mathds{P}}

\DeclareMathOperator*{\eqdef}{\stackrel{\triangle}{=}}

\def\BibTeX{{\rm B\kern-.05em{\sc i\kern-.025em b}\kern-.08em
    T\kern-.1667em\lower.7ex\hbox{E}\kern-.125emX}}

\begin{document}

\title{Robust Resource Sharing in Network Slicing \\ via Hypothesis Testing
\thanks{.}
}

\author{\IEEEauthorblockN{Panagiotis Nikolaidis and John Baras\\}
\IEEEauthorblockA{Department of Electrical \& Computer Engineering and the Institute for Systems Research\\
University of Maryland, College Park, MD 20742, USA\\
Email: \{nikolaid, baras\}@umd.edu}}

\maketitle

\begin{abstract}
In network slicing, the network operator needs to satisfy the service level agreements of multiple slices at the same time and on the same physical infrastructure. To do so with reduced provisioned resources, the operator may consider resource sharing mechanisms. However, each slice then becomes susceptible to traffic surges in other slices which degrades performance isolation. To maintain both high efficiency and high isolation, we propose the introduction of hypothesis testing in resource sharing. Our approach comprises two phases. In the trial phase, the operator obtains a stochastic model for each slice that describes its normal behavior, provisions resources and then signs the service level agreements. In the regular phase, whenever there is resource contention, hypothesis testing is conducted to check which slices follow their normal behavior. Slices that fail the test are excluded from resource sharing to protect the well-behaved ones. We test our approach on a mobile traffic dataset. Results show that our approach fortifies the service level agreements against unexpected traffic patterns and achieves high efficiency via resource sharing. Overall, our approach provides an appealing tradeoff between efficiency and isolation.
\end{abstract}


\section{Introduction}
\label{intro}
The need to support applications with diverse Quality of Service (QoS) requirements introduces several challenges in networking. A promising solution approach is to leverage network virtualization and software-defined networking to deploy multiple virtual networks on the same physical infrastructure. Each virtual network is then tailored to a specific application. 

For instance, a virtual network deployed for autonomous driving applications may drop packets whose age is larger than the sampling period at the vehicle to avoid congesting the network with outdated information. In contrast, virtual networks for file transfer may instead optimize throughput. Also, virtual networks used by private companies may require increased authentication functionality for security concerns.

The special case of customized virtual networks on the cellular infrastructure has attracted a lot of attention given that many applications require wireless connectivity. These virtual networks are often referred to as Network Slices (NSs).  They span the Radio Access Network (RAN), the Transport Network (TN) and the Core Network (CN) of the cellular infrastructure. Typically, the customer requesting a NS and the network operator sign a Service Level Agreement (SLA) that specifies the desired QoS delivered to the customer's traffic and the price paid to the operator. 

The fulfillment of multiple SLAs at the same time and on the same physical infrastructure is a difficult problem that operators need to solve. It is necessary that the operator provisions enough network resources in advance so that the network functions of each NS can deliver the promised QoS. Based on the provisioned resources, the operator can then compute the cost of the SLA and charge the customer.

Network functions that require provisioned resources span the whole cellular infrastructure. For example, in the RAN, the operator needs to provision Physical Resource Blocks (PRBs) for the Medium Access Control (MAC) scheduler of each NS. In the TN, routing paths need to be selected, and in the CN, processing units need to be reserved to execute various network functions such as the Access and Mobility Function (AMF) and the User Plane Function (UPF) of 5G systems.

An efficient provisioning approach is to enable resource sharing among NSs for statistical multiplexing gains. Resource sharing allows the deployment of an increased number of NSs on the same infrastructure which translates to lowered costs for the customers. Unfortunately, sharing may degrade the performance isolation of a NS. Indeed, unexpected traffic surges in a NS may result in resource contention. As a result, some NSs may not receive their fair share. This is highly undesirable since customers request a NS for premium service that should remain unaffected by the traffic in other NSs. 

On the other hand, the operator may provision resources exclusively for each NS which provides full isolation but at the cost of low resource efficiency. Motivated by this tradeoff, we propose the use of hypothesis testing to enhance performance isolation in resource sharing. Our approach consists of two phases, the trial phase and the regular phase.

In the trial phase, the traffic and the resources required by each NS are collected over a long period of time. This data is used as follows. First, a stochastic model is constructed for each NS that describes its normal traffic patterns. Specifically, a Markov Chain (MC) is used whose states describe the traffic and the resource demand in the NS. Second, the required provisioned resources are computed by estimating the percentiles of the resource demands of the NSs. All these quantities are computed based on Maximum Likelihood Estimation (MLE).

We note that the operator commits to fulfilling the SLA under the condition that the NS follows its normal behavior. This is a reasonable condition since the operator cannot provide QoS guarantees without any knowledge regarding the traffic of the NS. Indeed, the operator needs to know the traffic that the NS normally generates to compute the required provisioned resources and to charge the customer accordingly.

In the regular phase, the operator declines service to NSs that deviate from their normal behavior in case of resource contention. This is done to ensure that the well-behaved NSs receive their fair share of resources. The operator checks if such a deviation is present via hypothesis testing based on the Neyman-Pearson framework. Next, the operator splits the provisioned resources among the NSs that pass the test. Any remaining resources are then split among the rest of the NSs.

We apply our approach in the RAN where resource sharing is needed the most since the licensed spectrum is a scarce and expensive resource. Specifically, we wish to reduce the Physical Resource Blocks (PRBs) needed at the Base Station (BS) to satisfy the QoS requirements of all NSs. 

To test our approach, we use a dataset that contains real traffic as observed in base stations of a cellular network. The dataset is used to simulate the trial phase and derive the aforementioned stochastic models. Then, in the regular phase, we consider that some NSs follow traffic patterns that deviate from the previous models. Next, we analyze the effect of this excess traffic on the performance of the other NSs. We compare our approach to two baselines; exclusive bandwidth provisioning for each NS without resource sharing, and resource sharing without hypothesis testing. 

The results show that our approach enhances the robustness of the SLAs to anomalous traffic patterns while maintaining high efficiency via resource sharing. Overall, we provide evidence that resource sharing augmented by hypothesis testing strikes a good balance between efficiency and isolation. 
  
Our paper is structured as follows. In Sec. \ref{relwork}, we present the related literature and compare it with our work. In Sec. \ref{sysa}, we present the considered system architecture. In Sec. \ref{probform}, we formulate the overall goal of the system architecture as an optimization problem. In Sec. \ref{propsol}, we propose our solution approach. The simulation setup is described in Sec. \ref{ss} and the results in Sec. \ref{sr}. Lastly, Sec. \ref{concl} concludes the paper.

\section{Related Literature}
\label{relwork}
In \cite{imdea}, the authors studied the effect of overbooking strategies on resource allocation and service violations. Two schemes were investigated; perfect sharing and network slicing which correspond to resource sharing and exclusive resource reservation respectively. In perfect sharing, the BS sums the resource demands of all NSs and provisions $P^H$-percentile resources. In network slicing, performance isolation is considered by allocating exclusive $P^L$-percentile resources to each NS. The remaining resources needed to achieve $P^H-P^L$ fraction of time acceptance for each NS are computed based on past data. The authors provide insight regarding the tradeoff between resource efficiency and performance isolation via experimentation. However, no mechanisms to enhance isolation for the $P^H-P^L$ fraction of time are proposed.

The aforementioned tradeoff is also investigated in \cite{wiopt23}. The authors consider $P^L_i$-percentile resources assigned exclusively to each NS $i$. For the rest fraction of time $P^H_i-P^L_i$, each NS $i$ relies on resource sharing. The authors show that the multiplexing strategy that requires the least provisioned resources to satisfy the SLAs of all NSs is the Max-Weight scheduler if the resource demands follow MCs. However, no isolation mechanisms are provided in case of traffic anomalies.

A similar strategy is developed in \cite{dasilva}. The authors consider a fixed amount of provisioned resources and reserve some of it exclusively for each NS. The remaining resources are viewed as auxiliary and are dynamically provided to the NSs. The main focus of the paper is on developing overbooking strategies for the auxiliary resources with probabilistic guarantees. To do so, the authors assume that the operator has a model that forecasts the traffic of the NSs in the short term future. The NSs are then priced on a resource basis. Once again, this approach is susceptible to unexpected traffic patterns.

In \cite{zussman} and \cite{imdea-mobihoc}, a prediction model for the short-term demand of a NS is developed. The predictions are used to dynamically adapt the allocated resources for higher utilization. However, bandwidth adaptation alone cannot ensure the fulfillment of a SLA. Resources need to be provisioned so that the ones requested by the prediction model are available on short notice as often as the SLA requires.

The authors in \cite{survbanchs} conduct a survey on resource allocation schemes in network slicing. The schemes are coarsely divided into reservation-based and share-based. The authors conclude that the former provide higher isolation while the latter higher efficiency. Moreover, the schemes are also compared based on customizability, complexity, privacy and cost predictability.

A survey dedicated to the different interpretations of isolation is presented in \cite{isolation}. The authors discuss the concept of isolation in terms of performance, security, and dependability for various network functions in the RAN, TN, and CN. Here, we are primarily concerned with performance isolation.

A research area that is closely related to our approach is anomaly detection for which multiple methods have been suggested over the years \cite{anomaly1,anomaly2,anomaly3,anomaly4,quickchange}. Here, we use a statistical method since we consider hypothesis testing. We note that anomaly detection has been used before in network slicing \cite{nsanomaly1,nsanomaly2,nsanomaly3}. However in these works, the motivation behind its use is primarily security and data privacy. Here, we leverage anomaly detection to enhance isolation in resource sharing. 

Next, we note that similar issues regarding the fulfillment of SLAs while maintaining high resource utilization also appear in cloud computing. In \cite{salman}, the authors relate the resource allocation problem to a variation of an online knapsack problem. The authors then note that various other aspects need to be included such as SLA violation costs and resource migration delays that further complicate the problem. The authors in \cite{caglar} consider a neural network to predict user usage and dynamically allocate resources for higher utilization, an approach that is similar to \cite{zussman}.

Another related work in cloud computing is \cite{multitenant} where the tradeoff between performance isolation and fairness is addressed. The authors also consider the possibility of misbehaving customers with skewed and shifted demands. They propose the decomposition of the system-wide fair sharing problem into four smaller mechanisms and show robustness to customer misbehavior.

Overall, we believe that the related literature mostly considers a fixed set of provisioned resources that should be split among customers in a fair manner based on a utility maximization problem. However, a customer is interested in receiving the promised QoS as stated in the SLA. Hence, the fact that the customers receive their fair share according to utility maximization is of little value to them if the promised QoS is not delivered. Thus, resource provisioning and dynamic resource allocation must be studied jointly. Also, we believe that mechanisms to protect the SLAs from misbehaving traffic have not been investigated in detail. Here, we wish to address these two gaps in the literature by proposing provisioning mechanisms and hypothesis testing in resource sharing.

\section{System Architecture}
\label{sysa}
We consider $N$ NSs served by a single BS in the RAN. Each NS may have different type of QoS requirements. For instance, a NS may wish to upper bound the average packet delay of its users by a threshold, whereas another NS may wish to provide a constant target bitrate to each of its users. Thus, we consider that NSs may use different MAC schedulers. 

Let vector $W_i(t)$ denote the bandwidth demand of NS $i$ at slot $t$. Bandwidth demand $W_i(t)$ corresponds to the number of PRBs that the MAC scheduler of NS $i$ needs in order to provide the desired QoS throughout slot $t$ to NS $i$. 

To determine demand $W_i(t)$, the operator may need to observe the state of the NS $i$ at time $t$ denoted by $X_i(t)$. For instance, suppose that NS $i$ needs to deliver a constant target bitrate to each of its users. To do so, the operator first collects the Modulation and Coding Scheme (MCS) of each user at slot $t$, which composes the state $X_i(t)$ of the NS. Then, given the target bitrate, the operator may use Table 7.1.7.1-1 and Table 7.1.7.2.1-1 in 3GPP document \cite{3gpptables} to find the necessary number of PRBs $W_i(t)$. 

Note that for complex QoS requirements, the online determination of $W_i(t)$ based on $X_i(t)$ is not trivial. As a result, we consider a network function deployed at the BS which we call the Bandwidth Demand Estimator (BDE) whose objective is to compute $W_i(t)$ based on $X_i(t)$. Notice that the BDE enables bandwidth adaptation which is needed to achieve high efficiency in resource sharing. In \cite{BDE}, a BDE for packet delay requirements is developed based on a Reinforcement Learning (RL) algorithm and is then tested by experimentation on a 3GPP compliant cellular testbed. The design of new BDE is out of the scope of this paper.  

\begin{figure}
\centering
\includegraphics[width=\linewidth]{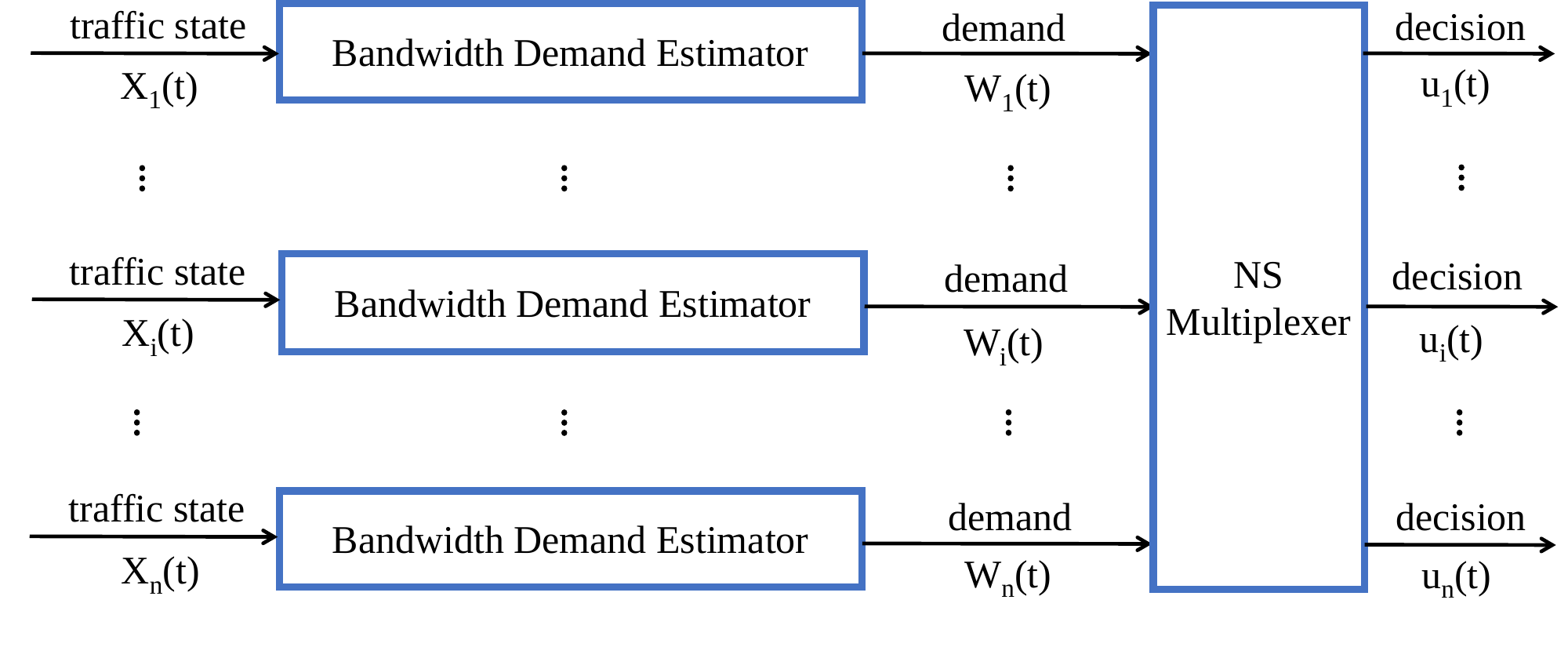}
\vspace{-0.6 cm}
\caption{We consider two network functions deployed at the BS. First, the BDE observes the state $X_i(t)$ of a NS and estimates the number of PRBs $W_i(t)$ needed to meet its desired QoS. Second, the NSM receives all bandwidth demands and decides which ones to satisfy given the limited bandwidth at the BS by producing binary decisions $u_i(t)$. The length of each slot $t$ depends on the timescales supported by these two network functions.}
\label{system}
\end{figure}

Next, since the provisioned PRBs at the BS may not suffice for all demands $\mathbf{W}(t) \: \eqdef \: (W_i(t))_{i \in [N]}$, the operator needs to decide which ones to accept. For this reason, we consider another network function at the BS which we call the Network Slice Multiplexer (NSM) that at each slot $t$ decides whether to allocate the $W_i(t)$ PRBs to the MAC scheduler of NS $i$. The NSM outputs a binary decision vector $\mathbf{u}(t)$ where $u_i(t)=1$ denotes acceptance of demand $W_i(t)$. Figure \ref{system} depicts the overall system architecture.

Notice that resource sharing is performed by the NSM. Therefore, in this paper, we focus only on the development of a NSM. We note that a preliminary solution was provided in \cite{wiopt23} where this system architecture was introduced. However, no mechanisms to protect SLAs against anomalies were considered and resource provisioning was not fully addressed.

\section{Problem Formulation}
\label{probform}
We consider that each customer $i$ states in the SLA that the operator needs to deliver the desired QoS for $P^H_i$ fraction of time. Such availability requirements are widely used in real networks for resource provisioning purposes, as in Google's software-defined network B4 \cite{after}. 

Notice that the previous statement is equivalent to the requirement that $u_i(t)=1$ for at least $P^H_i$ fraction of time. Let $T_s$ denote the number of slots during which all NSs are deployed and $W^c$ denote the provisioned bandwidth at the BS. Clearly, the operator wishes to satisfy all SLAs with minimum provisioned bandwidth. Thus, whenever a new NS needs to be deployed, the operator wishes to solve the following optimization problem:
\begin{align}
& \underset{W^c, \{\mathbf{u}(t)\}_{t \in [T_s]}}{\text{minimize}}  W^c \nonumber \\
& \text{s.t.:} \: \frac{1}{T_s}\sum\limits_{t=1}^{T_s}u_i(t) \geq P_i^H, \: \forall i \in [N], \nonumber \\ 
& \phantom{\text{s.t.:} \:} \mathbf{u}(t)^\top \mathbf{W}(t) \leq W^c, \: \forall t \in [T_s], \nonumber \\
& \phantom{\text{s.t.:} \:} \mathbf{u}(t) \in \{0,1\}^N, \: \forall t \in [T_s]. 
\label{overallprob}
\end{align}

The first constraint is the availability requirement as stated in the SLA. The second constraint states that the total accepted demand must be less than the provisioned resources $W^c$. The third constraint implies that the NS either receives the desired QoS or it does not. Next, note that (\ref{overallprob}) is a Mixed Integer Linear Programming (MILP) problem. Unfortunately, to solve it, the operator must know the future demands $\{\mathbf{W}(t)\}_{t \in T_s}$ and the exact system duration $T_s$ which is not possible.

\section{Proposed Solution Approach}
\label{propsol}
To approximately solve (\ref{overallprob}), we propose an approach with two phases; the trial phase and the regular phase. During the trial phase, a long sequence of traffic states and bandwidth demands $(X_i(t),W_i(t))$ is observed for each NS $i$. Based on these, a stochastic model is constructed for each NS that describes its normal behavior and the required provisioned bandwidth at the BS is estimated.

\begin{figure}
\centering
\includegraphics[scale = 0.3]{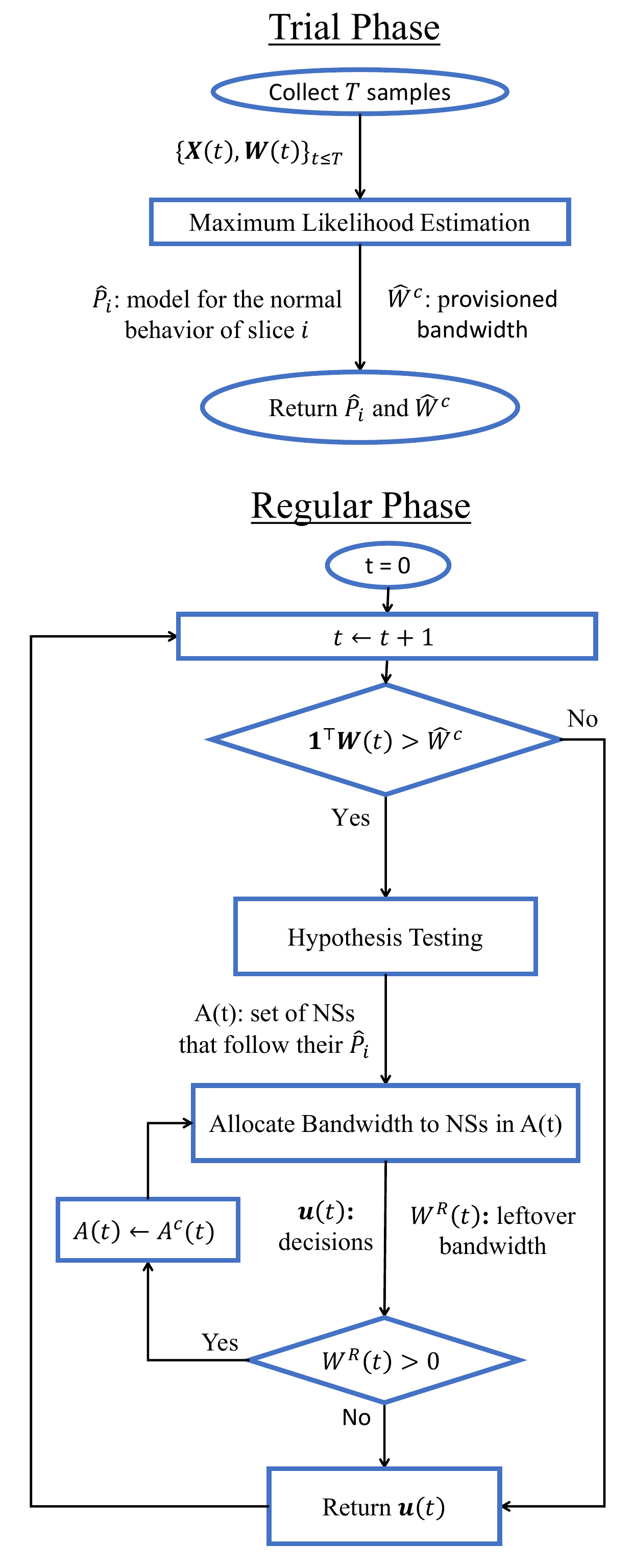}
\vspace{-0.3 cm}
\caption{When a new NS is deployed, the trial phase is initiated to obtain a stochastic model for the normal behavior of each NS and to estimate the required provisioned bandwidth. Next, the regular phase is initiated using the previously estimated provisions and models. Whenever there is resource contention, the NSM checks if each NS behaves as expected via hypothesis testing. Anomalous NSs are deprioritized in bandwidth allocation.}
\label{solafig}
\end{figure}

In the regular phase, the operator provisions the previously estimated bandwidth at the BS. Then, if the total bandwidth demand at some time exceeds the provisioned bandwidth, the NSM checks if the traffic generated by each NS was in accordance to its normal behavior as observed during the trial phase. The NSM performs this check by conducting a composite hypothesis test for each NS based on the Neyman-Pearson framework. The NSs that pass this test have prioritized access to the provisioned bandwidth. Any remaining bandwidth is then split among the other NSs. A flowchart of the overall solution approach is shown in Fig. \ref{solafig}.

\subsection{Trial Phase}
This phase is initiated when a new NS is deployed on the BS. To monitor the unknown resource demands of the new NS, the operator assigns as much bandwidth as possible to the BS to avoid situations where the poor QoS affects the behavior of the NS, e.g., users quitting due to large packet delays. The main goal during this phase is to describe the normal behavior of each NS by a stochastic model and estimate the provisioned bandwidth required at the BS to satisfy the SLAs of all NSs.

The stochastic model used greatly affects the estimation of the provisioned bandwidth. To gain insight on what model is appropriate, it is helpful to consider the models used by the BDE. For instance, in \cite{BDE}, the BDE is based on Reinforcement Learning (RL). In this case, the stochastic model for $(X_i(t),W_i(t))$ is a Markov Decision Process (MDP). Let $H_i(t) \eqdef \{(X_i(\tau),W_i(\tau))\}_{\tau=1}^{t-1}$ be the past state-action pairs. The MDP models the process $\{(X_i(t),W_i(t))\}_{t \in \mathds{N}}$ as follows:
\begin{align}
\prob(X_i(t+1)=x'| X_i(t)=x, W_i(t)=w, H_i(t)=h_i) \nonumber \\
= P_i(x'|x,w).
\label{rltran}
\end{align}

Suppose that the NS is deployed for a large amount of time, i.e., $T_s \to \infty$, and that the BDE wishes to minimize some total expected cost that depends on the state $X_i(t)$ and action $W_i(t)$. Then, it suffices to consider stationary policies \cite{bertsekas2} that allocate bandwidth $W_i(t)$ given state $X_i(t)$ as follows:
\begin{align}
\prob(W_i(t)=w| X_i(t)=x, H_i(t)=h_i) = \mu_i(w|x).
\label{rlpol}
\end{align}

From (\ref{rltran}) and (\ref{rlpol}), it is easy to see that the process $Z_i(t) \eqdef (X_i(t),W_i(t))$ follows a stationary MC:
\begin{align}
\prob(Z_i(t+1)=(x',w')| Z_i(t)=(x,w), H_i(t)=h_i) \nonumber \\
= P_i(x'|x,w)\mu_i(w'|x') \eqdef P_{Z_i}((x',w')|(x,w)).
\label{zitran}
\end{align}

Note that it is further reasonable to assume that the random processes $Z_i(t)$ and $Z_j(t)$ are conditionally independent for $j \neq i$ since information regarding other NSs $j$ do not provide any value to NS $i$ if $(X_i(t),W_i(t))$ are known. Let $\mathbf{X}(t) \: \eqdef \: (X_i(t))_{i \in [n]}$. Then, the vector $\mathbf{Z}(t)=(\mathbf{X}(t),\mathbf{W}(t))$ follows the MC:
\begin{align}
\prob(\mathbf{Z}(t+1)=\mathbf{z'}| \mathbf{Z}(t)=\mathbf{z}, \mathbf{H}(t)=\mathbf{h}) \nonumber \\
= \prod\limits_{i=1}^N P_{Z_i}(z_i'|z_i) \eqdef P_{\mathbf{Z}}(\mathbf{z'}|\mathbf{z}).
\label{ztran}
\end{align}

Hence, the stochastic model of interest is a MC. We further consider that all the above MCs are ergodic since state spaces $\mathcal{X}_i$ and action spaces $\mathcal{W}_i$ are finite. Also, each MC $Z_i(t)$ is expected to contains self-loops and to be possible to reach each state from all the other states. Therefore, each MC $Z_i(t)$ eventually converges to its unique stationary distribution $\pi_{Z_i}(z_i)$. Thus, the MC $\mathbf{Z}(t)$ has a unique stationary distribution $\pi_{\mathbf{Z}}(\mathbf{z})=\prod_{i=1}^N\pi_{Z_i}(z_i)$.

Since we wish to deliver the desired QoS to NS $i$ for at least $P^H_i$ fraction of time, let $W^H_i$ be defined as:
\begin{equation}
W^H_i \: \eqdef \: \arg\min_{w}\lim_{T_s \to \infty}\frac{1}{T_s}\sum_{t=1}^{T_s}\mathds{1}_{W_i(t) \leq w} \geq P^H_i.
\label{quant1}
\end{equation}
Let $\pi_{\mathbf{W}}(\mathbf{w}) \eqdef \sum_\mathbf{x}\pi_{\mathbf{Z}}(\mathbf{x},\mathbf{w})$ be the stationary probability of $\mathbf{W}(t)=\mathbf{w}$. Due to ergodicity, it follows that:
\begin{equation}
W^H_i = \arg\min_{w} \sum_{ \mathbf{w}:w_i \leq w}\pi_{\mathbf{W}}(\mathbf{w}) \geq P^H_i.
\label{quant2}
\end{equation}

Since the SLA dictates that the NS should be satisfied for $P^H_i$ fraction of time, we may ignore any bandwidth demand $W_i(t) >W^H_i$. Although treating demands higher than $W^H_i$ as $0$ is optimal for resource provisioning purposes, we consider it to be quite punishing. So instead we may transform all demands $W_i(t) >W^H_i$ as $W^H_i$, i.e., we may consider a transformed demand $g_i(W_i(t))=\min(W_i(t),W^H_i)$. Let $\mathbf{g}(\mathbf{W}(t)) \eqdef (g_i(W_i(t))_{i \in [N]}$ Thus, to satisfy all SLAs, the provisioned bandwidth $W^c$ should be:
\begin{equation}
W^c=\arg\min_{w}\lim_{T_s \to \infty}\frac{1}{T_s}\sum_{t=1}^{T_s}\mathds{1}_{\mathbf{1}^\top\mathbf{g}(\mathbf{W}(t)) \leq w} \geq \max_iP^H_i.
\label{provprob1}
\end{equation}
Similarly as before, due to ergodicity, (\ref{provprob1}) is equivalent to:
\begin{equation}
W^c=\arg\min_{s} \sum_{\mathbf{w}: \mathbf{1}^\top \mathbf{g(w)} \leq s}\pi_{\mathbf{W}}(\mathbf{w}) \geq \max_iP^H_i.
\label{provprob2}
\end{equation}

If the stationary distribution $\pi_{\mathbf{W}}(\mathbf{w})$ is known, then each $W^H_i$ can be found from (\ref{quant2}) via binary search. Then, functions $g_i$ are also known and $W^c$ can be computed from (\ref{provprob2}) also via binary search. Unfortunately, in practice we do not know the transition matrix of the MC $\mathbf{Z}(t)$ and therefore we cannot compute any of these quantities. As a result, we need to estimate the above quantities from the observed data. To estimate $W^H_i$ and $W^c$, we first estimate the stationary distribution $\pi_{\mathbf{w}}$ and proceed as described in (\ref{quant2}) and (\ref{provprob2}).

We use the Maximum Likelihood Estimator (MLE) of $\pi_{\mathbf{W}}(\mathbf{w})$ denoted by $\hat{\pi}_{\mathbf{W}}(\mathbf{w})$ which is simply the fraction of time that $\mathbf{W}(t) \leq \mathbf{w}$ in the observed sequence of length $T$ and can be easily updated online. We note that the above MLEs converge almost surely to the estimated quantities and the asymptotic rate of convergence is known due to asymptotic normality \cite{kay1}. 

Unfortunately, both these results are asymptotic and hold for large data sequences as $T \to \infty$. However, in practice, we are interested in the number of samples $T$ needed to obtain a specific accuracy of the estimated quantity with a certain probability. Hence, concentration inequalities such as the Chernoff-Hoeffding bounds are particularly useful here. Unfortunately, most of these bounds apply only to  Independent Identically Distributed (IID) random processes.

There are only few results that extend such bounds to MCs. For instance, the result in \cite{moulos} provides such a bound but requires the knowledge of a quantity that equals the largest expected time to transition from a state $x$ to a state $y$ for the first time among all $(x,y)$ pairs. Applying the result in \cite{moulos} for the MC $\mathbf{Z}(t)$, it follows that $\forall \epsilon > 0$:
\begin{align}
\prob\left(\frac{1}{T}\left|\sum_{t=1}^Tf(\mathbf{Z}(t))-\E[f(\mathbf{Z}(1))]\right| > \epsilon \right) \nonumber \\ 
\leq 2\text{exp}\left(-\frac{2T\epsilon^2}{(b-a)^2H_{\mathbf{Z}}^2}\right).
\label{moulos}
\end{align}
In the above, function $f$ maps states to a value in $[a,b]$ and $H_\mathbf{Z}=\max_{x,y}\E[T_y|\mathbf{Z}(1)=x]$ where $T_y=\inf \{ t \geq 0: \mathbf{Z}(t+1)=y\}$. The result in \cite{moulos} applies if the concerned MC has finite states, is irreducible and its initial distribution is one of its stationary distributions.

Note that $H_\mathbf{Z}$ is not known in practice since it depends on the transition matrix of the MC $\mathbf{Z}(t)$. Nonetheless, we argue that the periodic traffic patterns observed in real networks provide insight for its value. For instance, the period of the traffic patterns can be considered as an approximation of $H_{\mathbf{Z}}$.

To leverage (\ref{moulos}) for the estimation of $\pi_{\mathbf{w}}$, we consider $f_{\mathbf{w}}(\mathbf{Z(t)})=\mathds{1}_{\mathbf{W(t)=w}}$ and hence $a=0$ and $b=1$. Then, to bound by $\delta$ the probability that the absolute deviation of $|\hat{\pi}_{\mathbf{W}}(\mathbf{w})-\pi_{\mathbf{W}}(\mathbf{w})|$ exceeds $\epsilon$, the following sufficient condition is obtained from (\ref{moulos}): 
\begin{equation}
T \geq H_{\mathbf{Z}}^2\ln(2/\delta)/(2\epsilon^2).
\label{samples}
\end{equation}

Another Hoeffding bound for MCs is provided in \cite{princeton}. The authors show an optimal bound which requires the knowledge of the spectral gap $1-\lambda$ of its transition matrix, where $\lambda$ is its second largest eigenvalue in absolute value. Similarly as before, it follows from \cite{princeton} that it suffices to consider:
 \begin{equation}
T \geq \frac{1+\lambda}{1-\lambda}\ln(2/\delta)/(2\epsilon^2).
\label{samples2}
\end{equation}

It is known that $\lambda \in [0,1)$, thus the lower bound is minimized when $\lambda = 0$ which occurs when the rows of the transition matrix are equal \cite{princeton}. In this case (\ref{samples2}) coincides with the classical Hoeffding bound for the IID case. Thus, at best case scenario, the number of samples $T$ is $\ln(2/\delta)/(2\epsilon^2)$.

For some transition matrices obtained during experimentation, the lower-bound in (\ref{samples2}) was significantly lower than the one in (\ref{samples}). Thus, suppose that we consider $\epsilon=0.01$ and $\delta=0.01$ in $\ln(2/\delta)/(2\epsilon^2)$. We readily obtain $T \approx 120K$. Next, suppose that we observe a sample every $10$ seconds, i.e., the slot length in Fig. \ref{system} is 10 seconds. Then, the trial phase should be at least $2$ weeks. If we further wish to model every $8$-hour period in the day with a different MC, the duration of the trial phase needs to be at least $1.5$ month.

Unfortunately, we do not have a dataset with that many contiguous samples. However, we noticed during experimentation that even the best-case scenario bound may be loose since a smaller number of samples did not lead to SLA violations. Hence, we consider small $T$ values such as $7200$. Once the $T$ samples are obtained, the estimates of $W^H_i$ and $W^c$ are obtained from (\ref{quant2}) and (\ref{provprob2}) respectively via binary search.

Apart from the above estimates, we further need to estimate the transition matrix $P_{Z_i}$ of MC $Z_i(t)$ that describes the normal behavior of each NS $i$. The MLE for the transition probabilities $P_{Z_i}(z_i'|z_i)$ is simply the number of times there was a transition from state $z_i$ to state $z_i'$ over the total number of times state $z_i$ occurred. Once again, it is known that the MLE converges almost surely to the estimated quantity and that asymptotic normality holds \cite{kay1}. Unfortunately, we were not able to obtain any bounds for finite samples as previously even though \cite{moulos} provides a Chernoff-Hoeffding bound when the function $f$ receives two arguments.

The overall estimation procedure during the trial phase is summarized in Algorithm \ref{trialalgo}. We note that the algorithm can be simplified if certain conditions are met. First, given the independence assumption regarding the processes $Z_i(t)$, we may use the estimator $\hat{\pi}_{\mathbf{W}}(\mathbf{w})=\prod_{i=1}^N\hat{\pi}_{W_i}(w_i)$ and estimate the distribution of the total transformed demand $\mathbf{1}^\top\mathbf{\hat{g}}(\mathbf{w})$ by convolution. Even though we consider the processes $Z_i(t)$ to be independent, we do not follow the previous method and instead directly estimate the distribution of the whole demand vector $\pi_{\mathbf{W}}$ for simplicity. Next, note that if the demands are not transformed, i.e., $g_i(x)=x$, then we may skip many steps in Algorithm \ref{trialalgo} since we can directly estimate the cdf of $\mathbf{1}^\top \mathbf{w}$.

A procedure in Algorithm \ref{trialalgo} that requires a large amount of memory involves the estimation of the transition matrices $P_{Z_i}$. Note that each $Z_i(t)=(X_i(t),W_i(t))$ is composed by a vector of dimension $\text{dim}(\mathcal{X}_i)+1$ and the size of $\text{observed\_transitions}_i$ may be $|\mathcal{X}_i \times \mathcal{W}_i|^2$. However, notice that if the NSM knows the stationary policy $\mu_i(w|x)$ used by the BDE which is likely to be the case since both the BDE and the NSM run at the BS, then the NSM needs to estimate only the transition probabilities $P_i(x'|w,x)$ as in (\ref{zitran}). Furthermore, if the stationary policy used by the BDE is deterministic, i.e., $W_i(t)=\mu_i(X_i(t))$, then the NSM needs to estimate only the MC that its state $X_i(t)$ follows. Since $\epsilon$-soft polices are used in many popular RL algorithms and they can be approximated by a deterministic policy, we may estimate only the MC of process $X_i(t)$. Also, the estimation of the MC of $X_i(t)$ may be further simplified if its of each components follows an independent MC. Lastly, a crude approximation of the normal behavior of a NS may be obtained by assuming that each process $W_i(t)$ follows an independent MC.
\begin{algorithm}
$\mathbf{Input}$: parameters $\epsilon, \delta, H_\mathbf{Z}, P^H_i$\\
$\mathbf{Output}$: $\hat{W}^H_i,\hat{P}_{Z_i},\hat{W}^c$\\
$T \approx 7200$ \\
\tcc{ Collect  statistics online}
 \For{$t \leq T$}{
	get $\mathbf{Z}(t)=(\mathbf{X}(t),\mathbf{W}(t))$\\
	$\text{observed\_demands.add}(\mathbf{W}(t))$ \\ 
	counts($\mathbf{W}(t)$)+=1\\
	$\text{observed\_sums.add}(\mathbf{1}^\top \mathbf{W}(t))$\\
	$\text{total\_demand\_count}(\mathbf{1}^\top \mathbf{W}(t))+=1$\\
	\For{each NS $i$}{
		$\text{observed\_transitions}_i\text{.add}((Z_i(t-1)_,Z_i(t)))$\\
		$\text{transition\_counts}_i(Z_i(t-1),Z_i(t)) += 1$\\
		$\text{Z\_counts}_i(Z_i(t))+=1$\\

	}
	$\mathbf{Z(t-1)}=\mathbf{Z(t)}$

}
\tcc{MLE of pmf $\pi_\mathbf{w}$}
\For{$\mathbf{w} \in \text{observed\_demands}$}
{
	$\hat{\pi}_{\mathbf{W}}(\mathbf{w}) = \text{counts}(\mathbf{w})/T$\\
	$\text{counts}_i(w_i)+= \text{counts}(\mathbf{w})$\\
}
\tcc{MLE of $P^H_i$-percentiles $W^H_i$}
\For{each NS $i$}
{	
	sort $\mathcal{W}$ in increasing order\\
	\For{$w \in \mathcal{W}$}
	{
		$\hat{\pi}_{W_i}(w)=\text{counts}_i(w)/T$\\
		$ s+= \hat{\pi}_{W_i}(w)$\\
		\If{$s \geq P^H_i$}
		{
			$\hat{W}^H_i=s$\\
			break\\
		}
	}
	
}
\tcc{MLE of provisioned bandwidth $W^c$}
$\hat{g}_i(x)=\min(x,\hat{W}^H_i)$\\
\For{$\mathbf{w} \in \text{observed\_demands}$}
{
	$\text{observed\_gsums.add}( \mathbf{1}^\top \mathbf{\hat{g}}(\mathbf{w}))$\\
}
sort $\text{observed\_gsums}$ in increasing order\\
\For{$s \in \text{observed\_gsums}$}
{
	$\text{cdf} += \text{total\_demand\_count}(s)$\\
	\If{$\text{cdf} \geq \max_iP^H_i$}
		{
			$\hat{W}^c=s$\\
			break\\
		}
}
\tcc{MLE of transition matrices $P_{i}$}
\For{each NS $i$}
{	
	\For{$(z,z') \in \text{observed\_transitions}_i$}
	{
		$\hat{P}_{Z_i}(z'|z)=\text{transition\_counts}_i(z,z')/\text{Z\_counts}_i(z)$
	}
	
}

\caption{Estimation Procedure in Trial Phase}
\label{trialalgo}
\end{algorithm}
 
\subsection{Regular Phase}
At this point, each NS has been through the trial phase and a MC that describes its normal behavior has already been obtained. Also, the provisioned bandwidth should satisfy all NSs for $\max_iP^H_i$ fraction of time. Nonetheless, for the rest of the time, the bandwidth may not suffice. Moreover, a NS may generate unexpected traffic patterns, e.g, due to a special event, that increase its bandwidth demand. In such cases, the NSM needs to reject some of the demands.

Given that the provisioned bandwidth $\hat{W}^c$ was computed by considering that each NS follows its normal behavior as described by its MC $Z_i(t)$, it is fair to reject NSs that do not follow these MCs. For this reason, we consider that the NSM performs hypothesis testing when $\sum_{i=1}^NW_i(t)>\hat{W}^c$. The null hypothesis $H^0_i(t)$ is that NS $i$ follows so far the MC $\hat{P}_{Z_i}$ obtained during the trial period. The alternative hypothesis $H^1_i(t)$ is that the NS at some time $t-n+1$ switched to a different MC. Hence, a composite hypothesis test is considered. Selecting $H^0_i(t)$ implies that NS $i$ behaves normally so far and thus has prioritized access to the bandwidth. 

To conduct the hypothesis test, we consider the Neyman-Pearson framework \cite{kay2}. In this framework, we wish to bound the false alarm rate, i.e., the probability that we incorrectly select the alternative hypothesis $H^1_i(t)$, while maximizing the detection power, i.e., the probability that we correctly select $H^1_i(t)$. To do so, we check the likelihood ratio of the hypotheses and pick the alternative hypothesis if and only if the ratio is larger than a parameter $\gamma_i$. Let $Z_i(n,t)=\{Z_i(\tau)\}_{\tau=t-n+1}^t$ denote the last $n$ samples at time $t$. Then, the hypothesis $H^1_i(t)$ is selected if and only if:
\begin{equation}
L(Z_i(n,t)) = \frac{\prod\limits_{\tau=t-n+1}^{t}\hat{Q}_{Z_i}(Z_i(\tau)|Z_i(\tau-1))}{\prod\limits_{\tau=t-n+1}^{t}\hat{P}_{Z_i}(Z_i(\tau)|Z_i(\tau-1))} \geq \gamma_i,
\label{GLRT}
\end{equation}
where $\hat{Q}_{Z_i}$ is the MLE of the transition matrix of the MC that samples $Z_i(n,t)$ follow. 

The $\gamma_i$ in (\ref{GLRT}) that maximizes the detection power while bounding the false alarm rate by $\alpha_i$ is selected as follows:
\begin{equation}
\gamma_i = \arg\min_{\gamma}\prob(Z_i(n): L(Z_i(n)) > \gamma | H^0_i) \leq \alpha_i,
\label{pfa}
\end{equation}
where the subscript $t$ was dropped since the MC $Z_i(t)$ is stationary. The Neyman-Pearson Theorem \cite{kay2} states that this likelihood ratio test achieves the highest detection power given a bound on the false alarm rate out of all possible tests.

Note that ideally we wish to perform the hypothesis test $\forall n \leq t$ in order to check whether the transition matrix changed at any time in the past which corresponds to model change detection with unknown change time \cite{kay2}. However, such a test is too computationally intense to be conducted online. For this reason, we consider a fixed sample size $n$ and hence only check whether the MC changed at time $t-n+1$. 

Unfortunately, even for a fixed $n$, (\ref{pfa}) is quite complex to solve and therefore $\gamma_i$ cannot be computed exactly. Thus, we consider the asymptotic result that as $n \to \infty$, the random variable $2\ln L(Z_i(n))$ follows the chi-square distribution $\chi_r^2$ under hypothesis $H_i^0(t)$ \cite{kay2}. The distribution parameter $r$ is equal to the total degrees of freedom in the test.

In our case, the transition probabilities $\hat{Q}_{Z_i}(z'| z)$ are $|\mathcal{Z}_i|^2$ in total. However, for each $z$, it holds that $\sum_{z'}\hat{Q}_{Z_i}(z'|z)=1$. Since the transition matrix  $\hat{P}_{Z_i}$ is fixed, then $r=|\mathcal{Z}_i|^2 - |\mathcal{Z}_i|$. Thus, in the asymptotic regime, it follows from (\ref{pfa}):
\begin{equation}
\gamma_i = \text{exp}(F^{-1}_r(1-\alpha_i/2)),
\label{gamma1}
\end{equation}
where $F_r$ denotes the cdf of $\chi^2_r$ and $r=|\mathcal{Z}_i|^2 - |\mathcal{Z}_i|$. 

Since (\ref{gamma1}) holds only for large data records as $n \to \infty$, then we may not consider small $n$ to detect model changes that occurred in the recent past. Hence, if the behavior of the NS changes only for very brief periods of time and then reverts back to normal, we may never detect any anomalies. However, we argue that such fast and brief changes of time may not have a large effect on the resource allocation process. Lastly, note that the memory required for the test depends on $n$.

Let $A(t) \eqdef \{i \leq N: H_i^0(t) \text{ selected}\}$ denote the set of NSs that behave normally at time $t$ according to the hypothesis tests and let $B(t)$ denote the set of NSs that do not. We then split the resources among the NSs in $A(t)$ using the Max-Weight scheduler \cite{neely}. Specifically, we consider a deficit $d_i(t)$ owed to each NS $i$ that is defined as follows:
\begin{equation}
d_i(t+1) = [ d_i(t) -u_i(t)]^+ + P_i^H, \: \forall i.
\label{new-deficits}
\end{equation}
The bandwidth $\hat{W}^c$ is split among the NSs in $A(t)$ as follows:
\begin{align}
& \underset{\{u_i(t)\}_{i \in A(t)}}{\text{maximize}} \qquad \sum_{i \in A(t)}u_i(t)d_i(t) \nonumber\\
& \text{s.t.:} \: \sum\limits_{i: A(t)}u_i(t)W_i(t) \leq \hat{W}^c, \nonumber\\
& \phantom{\text{s.t.:} \:} u_i(t) \in \{0,1\}, \forall i \in A(t).
\label{new-max-weight}
\end{align}

The above scheduling procedure is motivated by the fact that if the deficits in (\ref{new-deficits}) are strongly stable, then the first constraint in (\ref{overallprob}) holds w.p.1 as $T_s \to \infty$ \cite{neely}. Although (\ref{new-max-weight}) is a binary knapsack problem and thus NP-Hard, we utilize the Google OR-Tools solver in \cite{ortools} to obtain an exact solution relatively fast using a branch and bound method. Lastly, we mention that this scheduler is also used in \cite{wiopt23}.

Let $u^*_i(t)$ for $i \in A(t)$ denote a solution of (\ref{new-max-weight}) and let $W^{R}(t) \eqdef \hat{W}^c-\sum_{i \in A(t)}u^*_i(t)W_i(t)$ denote the remaining bandwidth. Upon splitting bandwidth $\hat{W}^c$ as in (\ref{new-max-weight}), either all NSs in $A(t)$ are satisfied, i.e., $u^*_i(t)=1$ for all $i \in A(t)$ or there are some NSs in $A(t)$ whose demand $W_i(t)$ is not fully met. Let $A^R(t) \eqdef \{i \in A(t): u^*_i(t)=0\}$ denote the NSs in $A(t)$ whose demand was not fully met. 

Notice that if $A^R(t) \neq \emptyset$, then the remaining bandwidth $W^{R}(t) \eqdef \hat{W}^c-\sum_{i \in A(t)}u^*_i(t)W_i(t) \leq \min_{i \in A^R(t)}W_i(t)$. Otherwise, $\mathbf{u}^*(t)$ would not be an optimal solution to (\ref{new-max-weight}). To fully utilize the remaining bandwidth $W^R(t)$, we may solve a utility maximization problem for the NSs in $A^R(t)$. Let $W_i^R(t)$ denote the bandwidth received by NS $i$. Then, we may consider a simple utility function $U_i(W_i^R(t)) \eqdef W_i^R(t)/W_i(t)$ for each NS $i$, and maximize the total utility as follows:
\begin{align}
& \underset{\{W_i^R(t)\}_{i \in A^R(t)}}{\text{maximize}} \qquad \sum_{i \in A^R(t)}\frac{W_i^R(t)}{W_i(t)} \nonumber\\
& \text{s.t.:} \: \sum\limits_{i: A(t)}W^R_i(t) \leq W^R(t), \nonumber\\
& \phantom{\text{s.t.:} \:} 0 \leq W^R_i(t) \leq W_i(t), \forall i \in A^R(t), \nonumber\\
& \phantom{\text{s.t.:} \:} W^R_i(t) \in \mathcal{W}_i.
\label{utilmax}
\end{align}
Note that the solution to the above problem is to sequentially find the currently smallest $W_j(t)$ and assign as much bandwidth as possible to $W^R_j(t)$. Notice that if $\mathcal{W}_j=\mathds{R}$, then we simply assign the whole remaining bandwidth $W^R_j(t)=W^R(t)$, where $W_j(t) \leq W_i(t)$ $\forall i \in A^R(t)$, since $W^{R}(t) \leq \min_{i \in A^R(t)}W_i(t)$ as explained previously.

In case that $A^R(t) = \emptyset$, we may decide to split the remaining bandwidth $W^R(t)$ to the NSs in $B(t)$ that do not behave normally according to the hypothesis test. Thus, we may then solve (\ref{new-max-weight}) for the set $B(t)$ instead of the set $A(t)$. 

The overall detection and bandwidth allocation procedure during the regular phase is described in Algorithm \ref{regalgo}. Similarly as in the trial phase, the algorithm can be significantly simplified under certain conditions. For instance, if NSM knows the stationary policy $\mu_i(w|x)$ used by the BDE, then we may instead perform two simpler hypothesis tests; one that checks whether the stationary policy $\mu_i(w|x)$ is followed and one to check whether the transition probabilities $\hat{P}_i(x'|x,w)$ are followed as implied by (\ref{zitran}). These tests have smaller numbers of total degrees of freedom and can be performed in parallel.

Furthermore, in case the stationary policy is deterministic, i.e., $W_i(t)=\mu_i(X_i(t))$, then the NSM may simply first check if $W_i(\tau)=\mu_i(X_i(\tau))$, $\forall \tau$ in the observation window. If the condition does not hold, then the alternative hypothesis $H_i^1(t)$ is selected. Otherwise, the hypothesis test needs to only involve the parameters $\hat{Q}_i(x'|x, \mu_i(x))$. Similarly as before, if the state components of a NS follow independent MCs, then we may consider a simpler hypothesis test for each of them.
\begin{algorithm}
$\mathbf{Input}$:  $\hat{W}^H_i,\hat{P}_{Z_i},\hat{W}^c,n$\\
$\mathbf{Output}$: $\mathbf{u}(t),\mathbf{W}^R(t)$\\
set $r_i= |\mathcal{Z}_i|^2-|\mathcal{Z}_i|$ and $\gamma_i = \text{exp}(F^{-1}_{r_i}(1-\alpha_i/2))$\\
set $d_i(0)=P^H_i$\\
\For{$t \leq T_s$}
{
	set $\mathbf{u}(t) = \mathbf{0}$ and get $\mathbf{Z}(t)=(\mathbf{X}(t),\mathbf{W}(t))$\\
	\For{each NS $i$}{
		$n\_\text{samples}_i\text{.add}(Z_i(t))$\\
	}
		\If{$t>n$}
		{
			$n\_\text{samples}_\text{i}\text{.pop()}$
		}
		\If{$\mathbf{1}^\top\mathbf{W}(t) \leq \hat{W}^c$} 
			{
			$\mathbf{u}(t)=\mathbf{1}$
			}
		\Else
		{
			\For{each NS $i$}
			{
				an=$\mathbf{hypothesis\_ test}\text{(n\_{samples}}_i,\gamma_i,\hat{P}_{Z_i})$\\
				\If{an==False}
				{
					A(t).add(i)
				}
				\Else{B(t).add(i)}
			}
			$\mathbf{bandwidth\_allocation}()$\\
			\If{$A^R(t) = \emptyset$}
				{
					set $\hat{W}^c =W^R(t)$ and $A(t) = B(t)$\\			
					$\mathbf{bandwidth\_allocation}()$\\
				}
		}
	$d_i(t+1) = [ d_i(t) -u^*_i(t)]^+ + P_i^H$\\
}
\tcc{Functions}
\SetKwProg{Fn}{}{}{}
\Fn{$\mathbf{hypothesis\_test}(\text{n\_samples},\gamma_i, \hat{P}_{Z_i})$:}
{
	\For{$2 \leq k \leq n$}
	{
		
		$z' = \text{n\_samples[k]}$\\
		$z = \text{n\_samples[k-1]}$\\
		$\text{t\_counts}(z,z')+=1$\\
		$\text{Z\_counts}(z')+=1$
	}
	$L_0=L_1=1$\\
	\For{$2 \leq k \leq n$}
	{
		$z' = \text{n\_samples[k]}$, $z = \text{n\_samples[k-1]}$\\
		$\hat{Q}_{Z}(z'|z)=\text{t\_counts}(z,z')/ \text{Z\_counts}(z')$\\
		$L_1=\hat{Q}_{Z_i}(z'|z)L_1$, $L_0=\hat{P}_{Z_i}(z'|z)L_0$\\
	}
	\If{$L_1\geq \gamma_iL_0$}{return True}
	\Else{return False}
}
\Fn{$\mathbf{bandwidth\_allocation}()$:}
{
	solve Binary Knapsack Problem (\ref{new-max-weight})\\
			\For{each NS $i \in A(t)$}
			{
				\If{$u^*_i(t)==0$}
					{
						$A^R(t)$.add($i$)
					}		
			}
			\If{$W^R(t) > 0$}
				{
					solve linear optimization problem (\ref{utilmax})
				}
}
\caption{Resource Sharing in Regular Phase}
\label{regalgo}
\end{algorithm}

\section{Simulation Setup}
\label{ss}

\subsection{Schemes}
To properly evaluate our approach, we compare it to two other baselines. Overall, the three schemes are the following:

\textbf{Sharing and Testing (ShT):} This scheme is the proposed solution approach in Sec. \ref{propsol}. In this scheme, resource sharing is augmented by hypothesis testing to enhance isolation. By ShT$n$, we refer to this scheme run with sample size $n$ in hypothesis testing.

\textbf{Sharing (Sh):} In the trial phase, this scheme is identical to the previous one. In the regular phase ,the scheme skips hypothesis testing and sets $A(t)=[N]$ in Algorithm \ref{regalgo}.

\textbf{No Sharing (NoSh):} This scheme provisions $\hat{W}^H_i$ bandwidth for each NS $i$ which is the estimation of the $P^H_i$-percentile of process $W_i(t)$ as obtained in the trial phase. Clearly, the total provisioned bandwidth is $\sum_i\hat{W}^H_i$. In the regular phase, the demand $W_i(t)$ is guaranteed to be accepted if $W_i(t) \leq \hat{W}_i^H$. Any leftover bandwidth is split among the other NSs. The scheme is run by skipping hypothesis testing and considering $i \in A(t)$ iff $W_i(t) \leq \hat{W}_i^H$ in Algorithm \ref{regalgo}.

\subsection{Metrics}
The schemes are compared based on two metrics. The first one is the required provisioned bandwidth $\hat{W}^c$ as estimated during the trial phase. The second one is the resulting acceptance ratio in the regular phase denoted by $a_i \eqdef \sum_{t=1}^{T_s}u_i(t)/T_s$. Notice that $a_i$ should be at least $P_i^H$ as shown in the first constraint of optimization problem (\ref{overallprob}).

Moreover, to gain more insight regarding the performance of each scheme, we also plot the ratio of correct and wrong rejections of each NS $i$ denoted by $r^c_i$ and $r^w_i$ respectively. The first ratio is the number of times that hypothesis testing rejected NS $i$ and it was indeed anomalous at that time over the number of times that hypothesis testing occurred and NS $i$ was anomalous. The second quantity measures the ratio of wrong rejections similarly.

Notice that $r^c_i$ and $r^w_i$ seem similar to the power and false alarm rate of the detector used in hypothesis testing. However, they differ from them since the detector checks if all $n$ previous samples where generated by the expected stochastic model, not just the most recent one. Hence, if a NS $i$ stops misbehaving at $t$, the detector is designed to reject it at time $t=t+n/2$ even if NS $i$ is not anomalous at that time. Therefore, the quantities $r^c_i$ and $r^w_i$ allow us to investigate the effect of sample size $n$ on performance during transition periods where a NS starts or stops misbehaving.

\subsection{Dataset}
To evaluate the aforementioned schemes, we use the dataset in \cite{dataset} which was made publicly available by the IMDEA Networks Institute\footnote{\label{note1}https://git2.networks.imdea.org/wng/madrid-lte-dataset}. It contains LTE mobile traffic at several BSs in Spain that was captured around 2020. The authors obtained these measurements by running a passive monitoring tool called Falcon \cite{falcon} on a Linux laptop that was connected to a USRP B210 Software Defined Radio (SDR). The authors then connected to various BSs and used Falcon to decode the Physical Downlink Control Channel (PDCCH) sent by the BS to the connected  User Equipments (UEs) in order to extract resource allocation information with millisecond granularity.

\begin{figure}
\centering
\includegraphics[width=\linewidth]{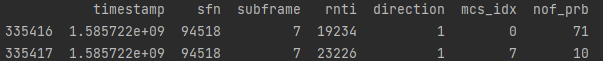}
\vspace{-0.6 cm}
\caption{A small sample of a parquet file found in the dataset. Here, two lines exist with the same timestamp and $\text{direction}=1$. This indicates that two UEs were receiving traffic from the BS during that subframe.}
\label{excerpt}
\end{figure}

The obtained raw data is stored as parquet files. Each line in the files contains the following: the unix timestamp, the system frame and the subframe number in LTE, the Radio Network Temporary Identifier (RNTI) of the UE, the direction of the traffic, i.e., uplink or downlink, the MCS index used by the UE and the number of PRBs utilized by the UE at that particular subframe. A sample of the file is shown in Fig. \ref{excerpt}.

\subsection{Extraction of state time series $\{X_i(t)\}_{t \in \mathds{N}}$} 
Since the system architecture proposed in Fig. \ref{system} is not deployed at the BSs, we use the dataset to create a time series for $\mathbf{Z}(t)$ for the evaluation of our proposed approach. We primarily focus on downlink traffic since it comprises most of the mobile traffic. Hence, we wish to obtain a time series for the states $X_i(t)$ and their corresponding bandwidth demands $W_i(t)$ for each NS $i$ in downlink.

To do so, we consider that the state $X_i(t)$ includes the number of UEs in Radio Resource Control (RRC) Connected state and their average MCS at time $t$. Notice that the number of RRC connected UEs at some time $t$ differs from the UEs that are actively transmitting or receiving at time $t$. This is the case since the MAC scheduler may wish to prioritize some UEs and allocate the whole bandwidth to them. Therefore, although some UEs may have data to transmit or receive, the MAC scheduler may not schedule them for transmission. As a result, we cannot simply count the number of lines in the data file at some time $t$ as in Fig. \ref{excerpt} to approximate the number of UEs in RRC Connected state.

Instead, we utilize the approach as in \cite{dataset} where the authors use a method that maps the temporary RNTIs to UE identifiers. This method was originally developed in \cite{rnti} and it essentially attempts to find the expiration period of the temporary RNTIs. Based on this estimation, the number of UEs in RRC  Connected state can be estimated since an RNTI acts as a UE identifier within its expiration period. The authors applied the method in \cite{rnti} on their raw data files and obtained an estimation of the number of RRC Connected UEs with one second granularity which is also included in their final dataset. Hence, for the first component of the state $X_i(t)$ we use this readily available data.

Regarding the second component of the state $X_i(t)$, we consider the average MCS over all users actively transmitting at time $t$. We compute this simply by taking the average of the "mcs\_idx" column in Fig. \ref{excerpt} for lines with $\text{timestamp} = t$ and $\text{direction} = 1$ since we are interested in downlink traffic.

\subsection{Definition of desired QoS}
In order to derive the time series for $Z_i(t)$ based on the times series $X_i(t)$, it is necessary to define the desired QoS of each NS $i$. We consider a simple QoS requirement that facilitates an easy map from $X_i(t)$ to $W_i(t)$. Specifically, we consider each NS $i$ requires a constant bitrate $R_i$ for each UE that is in RRC Connected state at each slot $t$.

\subsection{Extraction of bandwidth demand time series $\{W_i(t)\}_{t \in \mathds{N}}$}
We compute the required PRBs $W_i(t)$ to provide bitrate $R_i$ to each connected UE at time $t$ based on Table 7.1.7.1-1 and Table 7.1.7.2.1-1 in 3GPP document \cite{3gpptables} as mentioned previously in Sec. \ref{sysa}. Notice that the second component of $X_i(t)$, i.e., the average MCS over all users at time $t$ and the bitrate $R_i$ are the two elements needed to obtain the PRBs needed for a single UE based on the previous tables. Then, we multiple this amount of PRBs by the first component of $X_i(t)$, i.e., the number of connected users, to compute the bandwidth demand $W_i(t)$. Lastly, we assume that all UEs support 2x2 and thus we multiple the contents of the second table by $2$.

We note that it would be more accurate to repeat this process for each UE individually. However, this would require that $X_i(t)$ contains the MCS index of each UE which requires large memory. Although this is feasible, we consider the average MCS index over all UEs  and then multiple the resulting PRBs by the number of UEs for simplicity.

We mention that an alternative way to derive the bandwidth demand at time $t$ would be to use the raw data files by summing the "nof\_prb" column for all entries with $\text{timestamp} = t$ and $\text{direction} = 1$. In this case, the bandwidth demand $W_i(t)$ is approximated by the total number of PRBs allocated to the NS at time $t$. This is reasonable assuming that the BS delivered the desired QoS to the UEs.

Although this is a simple approach, we noticed that the total PRBs for some subframes $t$ exceeded $100$ PRBs which cannot be true since the cell bandwidth at the BS is $20$ MHz. Upon contacting the authors in \cite{dataset}, we were informed that such anomalies happen when the PDCCH is decoded erroneously due to low Signal to Noise Ratio (SNR) at the SDR. For this reason, we consider the method described previously to derive the bandwidth demands $W_i(t)$ over time. We note that we could not find any other publicly available dataset that contains granular resource allocation information at the BS.

Lastly, notice that the above procedure creates a deterministic map $W_i(t)=\mu_i(X_i(t))$ and hence it suffices to consider only the MC of $X_i(t)$ when conducting hypothesis testing as explained in the last paragraph of Sec. \ref{propsol}.

\subsection{Time, State and Action Aggregation}
With the previous procedures, we obtain a time series $Z_i(t)$ with one second granularity. However, it may not be realistic that the BDE operates in such a fast time scale. For this reason, we create a new time series $Z_i(t)$ by aggregating every $D$ seconds where $D$ corresponds to the slot length in Fig. \ref{system}. Here, we typically consider $D=10$ seconds. The time aggregation of the state series $X_i(t)$ is conducted by sampling the original series every $D$ entries.  In general, the aggregation of the $W_i(t)$ series is performed by representing every $D$ entries by their maximum to approximate the output of a BDE based on RL that operates every $D$ seconds and learns a map between $X_i(t)$ and $W_i(t)$. However, here we aggregate demands in time by $W_i(t)=\mu_i(X_i(t))$ as previously.

We also consider another form of aggregation. Given that the BDE may run a RL algorithm, we consider aggregation in the state and action space. For instance, we may consider that the first component of $X_i(t)$ takes values that are multiples of a constant $U_i=10$. Thus, if the actual number of UEs at time $t$ is between $[10,20)$, then $X_i(t)=10$. Such aggregations in the state and action space may significantly reduce the convergence time and the memory requirements in the BDE with little performance loss. Thus, we consider such aggregation constants for each NS $i$ denoted by $U_i$, $M_i$ and $W_i$ for the first state component, the second state component and the action respectively.

\subsection{Extraction of time series $\{\mathbf{Z}(t)\}_{t \in \mathds{N}}$}
So far we described how to extract a single time series $Z_i(t)$. However, we need to extract multiple such series to compose $\mathbf{Z}(t)$ when considering multiple NSs. To this end, each time series $Z_i(t)$ should be extracted from traffic data from the same BS and time period to consider traffic that competes for the same resources. This data should then be split into chunks by associating different groups of UEs to different NSs.

Unfortunately, this is not feasible since the dataset does not contain unique UE identifiers as mentioned earlier. Thus, we should then consider data samples that may originate from different BSs but still correspond to the same time period. However, we could not find any such data samples. A possible explanation is that the authors in \cite{dataset} used only one SDR for their measurements and thus could not obtain data from different BSs at the same time. As a result, we collect data samples from different BSs and time periods which we then associate to different NSs. Nonetheless, the time periods considered have certain common characteristics. 

Specifically, we consider data collected between specific hours, e.g., from 17:00 to 22:00 every day from Monday to Friday. Then, the samples collected from Monday to Thursday are used for the trial phase and the ones collected on Friday are used for the regular phase. The previous example results to $T=7200$ samples. We note that the selection of a certain range of hours is motivated by the fact that real traffic varies considerably throughout the day. Thus, we consider that a single MC $P_{Z_i}$ cannot model the traffic during the whole day but only during specific hours of each day.

Lastly, we note that the measurement in the raw data files are sometimes sparse. Therefore, the specific hours considered vary from experiment to experiment in order to obtain data samples that are dense. As a result, we do not utilize (\ref{samples}) to determine the number of samples needed during the trial phase. Instead, we obtain the data as described previously.

\subsection{Creation of anomalies}
As mentioned in Sec. \ref{propsol}, the anomalies considered are changes in the transition matrix of the process $Z_i(t)$. In case $W_i(t)= \mu_i(X_i(t))$, then it suffices to consider variations for the $P_{X_i}$ transition matrix. Moreover, if the state components follow independent MCs, then it suffices to consider variations to one of these MCs. As a result, we primarily modify the transition matrix $P_{U_i}$ that the number of connected users $U_i(t)$ follows. Such modifications may model an increase in the number of connected UEs to the BS.

Let $P'_{U_i}$ denote the new transition matrix followed by $U_i(t)$ and let $P'_{U_i}(u'|u)$ denote its elements. To properly evaluate our approach, the new matrix $P'_{U_i}$ must be constructed carefully. Otherwise, the anomalies may be easily detected by the hypothesis testing procedure. For instance, the matrix $P'_{U_i}$ should not contain any new entries, i.e., $P'_{U_i}(u'|u) = 0$ if the old entry $P_{U_i}(u'|u) = 0$. In addition, we are interested in new matrices $P'_{U_i}$ that produce higher demands $\mu_i(X_i(t))$ and result in resource contention that may negatively impact the performance of the other NSs. 

\begin{figure}
\centering
\includegraphics[width=\linewidth]{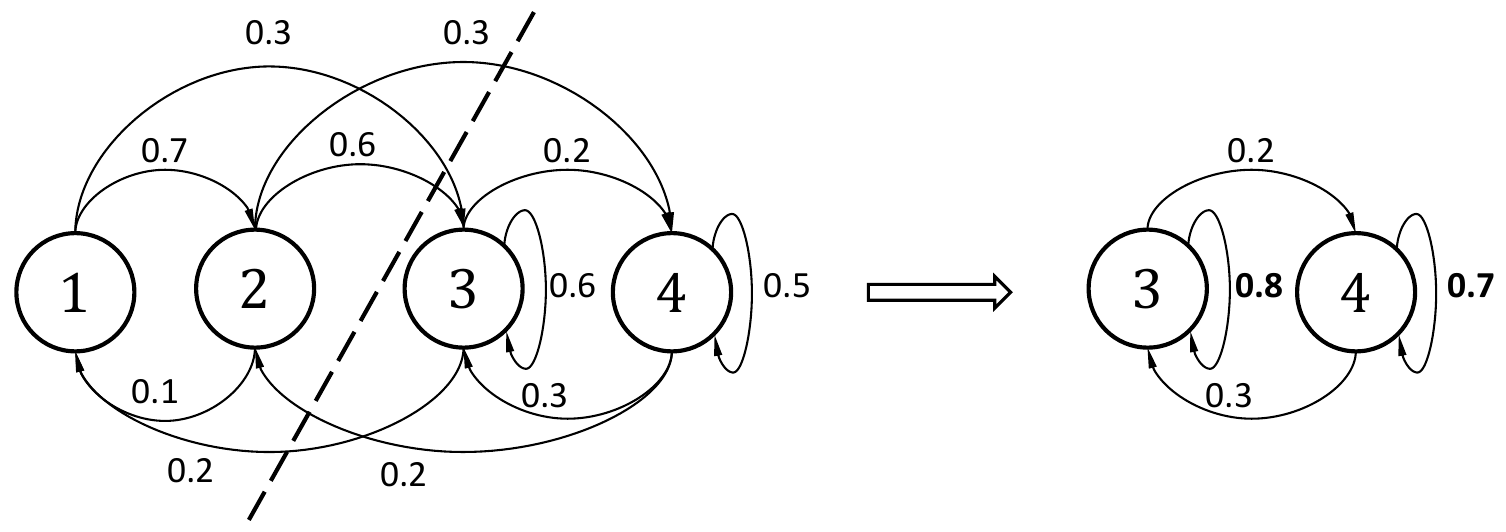}
\vspace{-0.7 cm}
\caption{Anomalies are created by changing the MC of a NS. Here, we remove the lowest $\beta=50\%$ states of the MC. Then,  each state's largest transition probability is increased so that each row of the new matrix sums to 1.}
\label{new_trans}
\end{figure}

Due to the above, we construct the new matrix $P'_{U_i}$ by deleting the lowest $\beta$\% of states in the MC of $U_i(t)$. To do so without creating new entries, the transition probability from a state $s$ to a removed state $s'$ is added to the largest transition probability from $s$. As a result, we create a new matrix that is similar to the original one but whose states correspond to higher number of users. Thus, the new matrix skews the behavior of the NS towards generating higher bandwidth demands. Figure \ref{new_trans} illustrates the previous procedure.

Once the new matrix $P'_{U_i}$ is obtained, an anomalous sequence $\{U_i(t)\}_{t_s \leq t\ \leq t_e}$ can be generated where $t_s$ and $t_e$ denote its start and end respectively. Notice that $t_s$ must be chosen carefully so that there is a smooth transition between the old and new sequence. Otherwise, the anomaly is easy to detect. For this reason, we choose $t_s$ as the earliest time that the old sequence arrives at one of the new states in $P'_{U_i}$. However, we also wish that the anomaly starts once the sample size $n$ is complete to consider the effect of old samples in the detection. Thus, we impose that $t_s > n$. For similar reasons, we wish that $t_e < T_s - n$. Lastly, we set $t_e$ as high as possible given the previous constraints so that the anomaly can have an effect on resource allocation and create resource contention.

\section{Simulation Results}
\label{sr}

\subsection{Test Scenario 1}
In the first test scenario, we consider two NSs. Each NS needs to constantly deliver $R_0=R_1=1$ Mbps to each connected UE for at least $P^H_0=P^H_1=0.9$ fraction of slots. The data for NS $0$ and NS $1$ are taken from the "I-1815-raw-df-ms.parquet" and "I-2650-raw-df.ms.parquet" files respectively. The data collection period for both NSs is from 17:00 to 22:00 each day from Monday, May 25 to Friday, May 29, 2020. The days from Monday to Thursday comprise the trial phase, whereas Friday comprises the regular phase. 

\begin{figure}
\centering
\includegraphics[width=\linewidth]{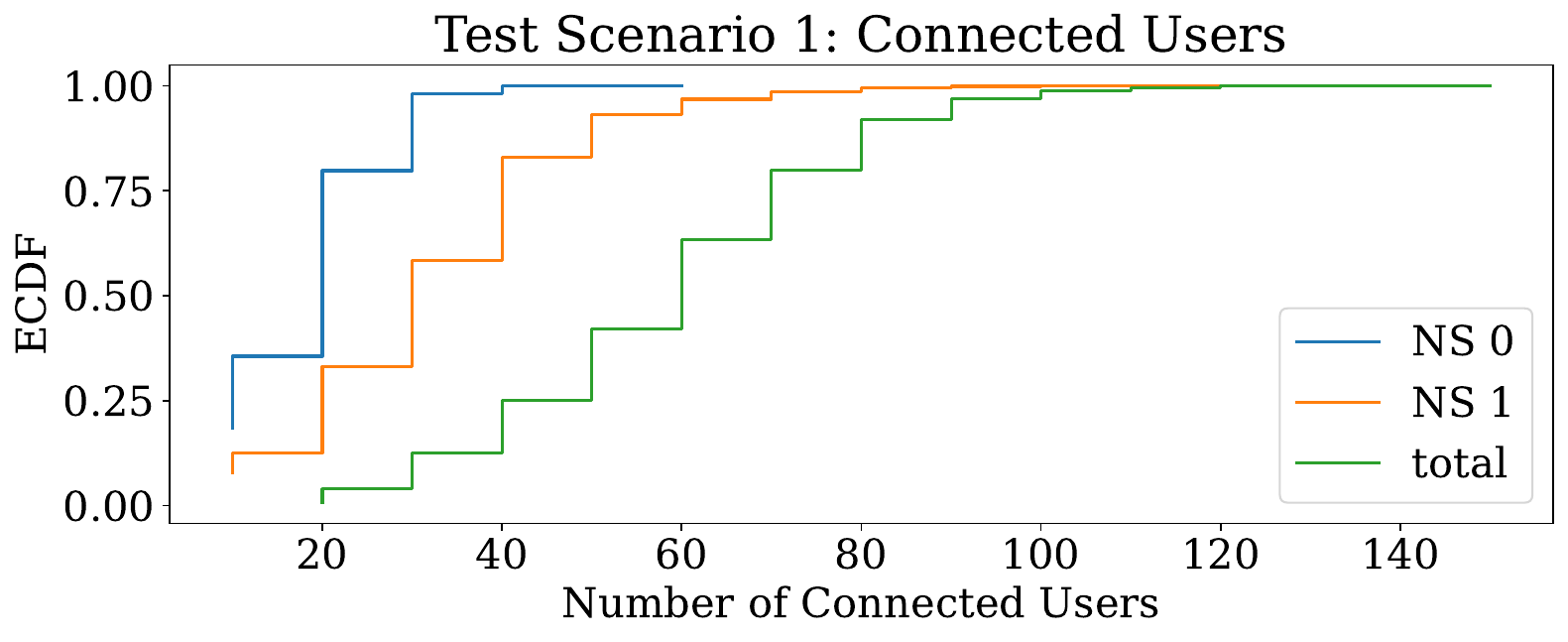}
\vspace{-0.8 cm}
\caption{The ECDF of the connected users in the regular phase.}
\label{ts1_user}
\end{figure}

\begin{figure}
\centering
\includegraphics[width=\linewidth]{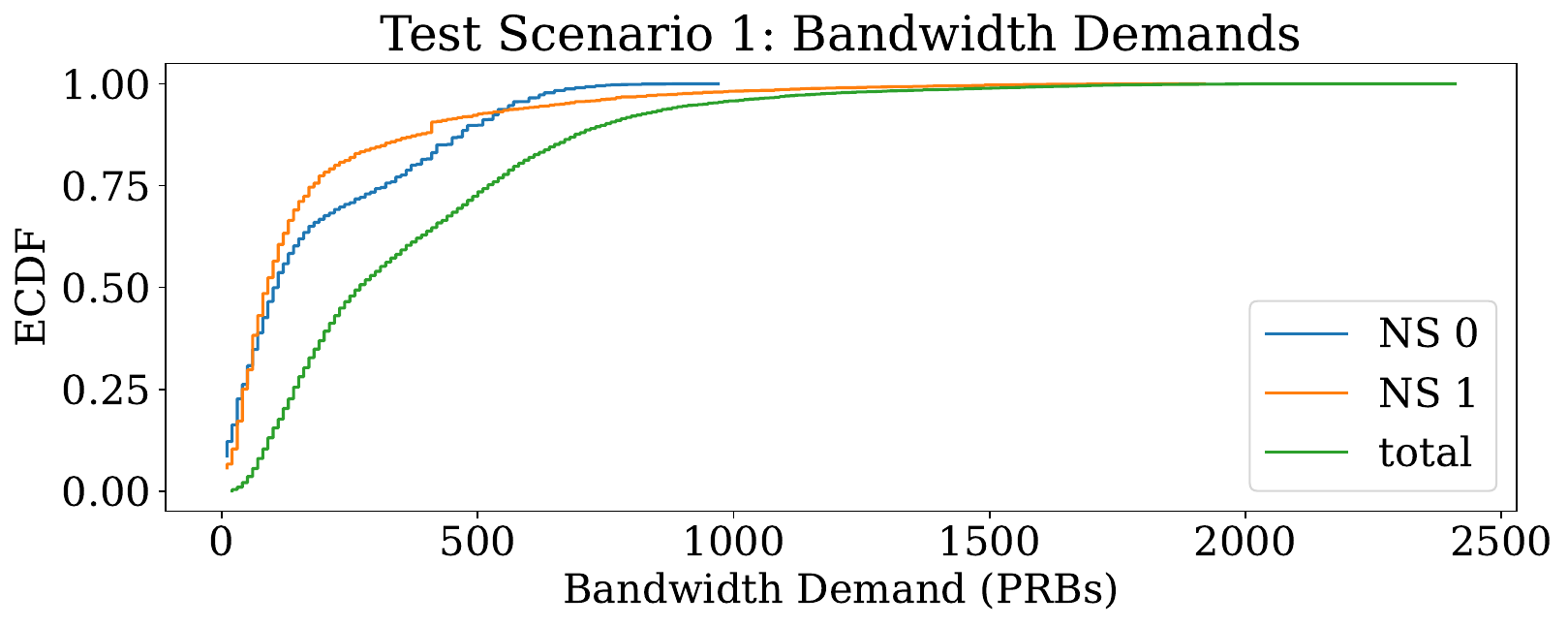}
\vspace{-0.8 cm}
\caption{The ECDF of the bandwidth demand in the regular phase.}
\label{ts1_PRB}
\end{figure}

We first consider that all NSs behave normally during the regular phase to obtain a reference point. Then, we consider that NS $0$ is anomalous for various $\beta$. We plot the metrics mentioned previously for each scheme and for each case. We vary the sample size $n$ to determine its effect on performance. We depict some traffic statistics of this test scenario in Fig. \ref{ts1_user} and Fig. \ref{ts1_PRB}. The results are shown in Tables \ref{table1}-\ref{table4}.

In Table \ref{table1}, we verify that the provisioned bandwidths estimated in the trial phase suffice for both NSs and the SLAs are fulfilled. Notice that $a_0$ and $a_1$ in the regular phase are slightly higher than the target $P^H_0=P^H_1=0.9$. This may indicate a small discrepancy between the statistics of the data in the trial phase and in the regular phase. However, the increase is probably due to the fact that the no sharing scheme allows the use of the idle $W^H_i$ of NS $i$ by other NSs $j$.

In Table \ref{table2}, NS $0$ is anomalous with $\beta=0.5$. As expected, the no sharing scheme protects the performance of NS $1$. However, in the sharing scheme, the SLA of NS $1$ is violated since $a_1 < 0.9$. In contrast, the sharing and testing scheme protects the SLA of NS $1$ when $n \geq 100$. This is also reflected in the fair rejection ratio where $r^c_0$ becomes high at $n=100$. This indicates that hypothesis testing starts to detect anomalies for $n \geq 100$. Also, notice that $r^w_0$ also increases as $n$ does since more outdated samples are stored which is detrimental when an anomaly stops. Since the anomaly exists almost throughout the simulation, these detrimental effects are particularly pronounced here. However, these effects would wash off in longer simulations. Also, we note that the above issue affects only NSs that at some point were anomalous for which the operator is not required to fulfill their SLA. 

Similar observations can be made for the cases where $\beta=4/6$ and $\beta=5/6$ in Tables \ref{table3} and \ref{table4} respectively. Here, also note that anomalies can be detected for a smaller sample size $n=50$ since they are more pronounced and thus the detector requires fewer samples to accurately detect them. Overall, in the cases where the simple sharing scheme fails, the proposed scheme satisfies the SLAs with $19\%$ less bandwidth than the no sharing scheme.

\begin{table}
\centering
\small
\caption{Test Scenario 1: Results when all NSs behave normally.}
\vspace{-2mm}
\begin{tabular}{lccc}
\hline
 Scheme   &    PRBs &   $a_0$ (\%) &   $a_1$ (\%) \\
\hline
 NoSh     &     940 &           98 &           95 \\
 Sh       &     760 &           94 &           93 \\
 ShT50    &     760 &           94 &           93 \\
 ShT100   &     760 &           94 &           93 \\
 ShT150   &     760 &           94 &           93 \\
 ShT200   &     760 &           94 &           93 \\
 ShT250   &     760 &           94 &           93 \\
\hline
\end{tabular}
\label{table1}
\end{table}

\begin{table}
\centering
\small
\caption{Test Scenario 1: Results when NS $0$ is anomalous with $\beta=0.5$.}
\vspace{-2mm}
\begin{tabular}{lrrrll}
\hline
 Scheme   &    PRBs &   $a_0$ (\%) &   $a_1$ (\%) & $r^c_0$ (\%)   & $r^w_0$ (\%)   \\
\hline
 NoSh     &     940 &           91 &           95 & -              & -              \\
 Sh       &     760 &           84 &           83 & -              & -              \\
 ShT50    &     760 &           84 &           83 & 0              & 0              \\
 ShT100   &     760 &           69 &           95 & 97             & 2              \\
 ShT150   &     760 &           69 &           95 & 98             & 12             \\
 ShT200   &     760 &           69 &           95 & 98             & 26             \\
 ShT250   &     760 &           68 &           95 & 98             & 40             \\
\hline
\end{tabular}
\label{table2}
\end{table}

\begin{table}
\centering
\small
\caption{Test Scenario 1: Results when NS $0$ is anomalous with $\beta=0.67$.}
\vspace{-2mm}
\begin{tabular}{lccccc}
\hline
 Scheme   &    PRBs &   $a_0$ (\%) &   $a_1$ (\%) & $r^c_0$ (\%)   & $r^w_0$ (\%)   \\
\hline
 NoSh     &     940 &           73 &           94 & -              & -              \\
 Sh       &     760 &           73 &           89 & -              & -              \\
 ShT50    &     760 &           69 &           95 & 99             & 2              \\
 ShT100   &     760 &           69 &           95 & 99             & 12             \\
 ShT150   &     760 &           69 &           95 & 99             & 25             \\
 ShT200   &     760 &           69 &           95 & 99             & 36             \\
 ShT250   &     760 &           69 &           95 & 99             & 40             \\
\hline
\end{tabular}
\label{table3}
\end{table}

\begin{table}
\centering
\small
\caption{Test Scenario 1: Results when NS $0$ is anomalous with $\beta=0.83$.}
\vspace{-2mm}
\begin{tabular}{lrrrll}
\hline
 Scheme   &    PRBs &   $a_0$ (\%) &   $a_1$ (\%) & $r^c_0$ (\%)   & $r^w_0$ (\%)   \\
\hline
 NoSh     &     940 &           73 &           94 & -              & -              \\
 Sh       &     760 &           73 &           89 & -              & -              \\
 ShT50    &     760 &           69 &           95 & 99             & 2              \\
 ShT100   &     760 &           69 &           95 & 99             & 12             \\
 ShT150   &     760 &           69 &           95 & 99             & 25             \\
 ShT200   &     760 &           69 &           95 & 99             & 36             \\
 ShT250   &     760 &           69 &           95 & 99             & 40             \\
\hline
\end{tabular}
\label{table4}
\end{table}

\subsection{Test Scenario 2}
In the second test scenario, we consider again two NSs. We consider required constant bitrates $R_0=1$ and $R_1=2$ Mbps and QoS delivery for $P^H_0=P^H_1=0.9$ fraction of slots. The data for NS $0$ and NS $1$ are taken from the "I-1796-raw-df-ms.parquet" and "I-1815-raw-df.ms.parquet" files respectively. The data collection period for NS $0$ is from 14:00 to 19:00 each day from June 17 to June 19, 2020 and for NS $1$ from 14:00 to 19:00 each day from May 25 to May 27, 2020. The last day corresponds to the regular phase.

\begin{figure}
\centering
\includegraphics[width=\linewidth]{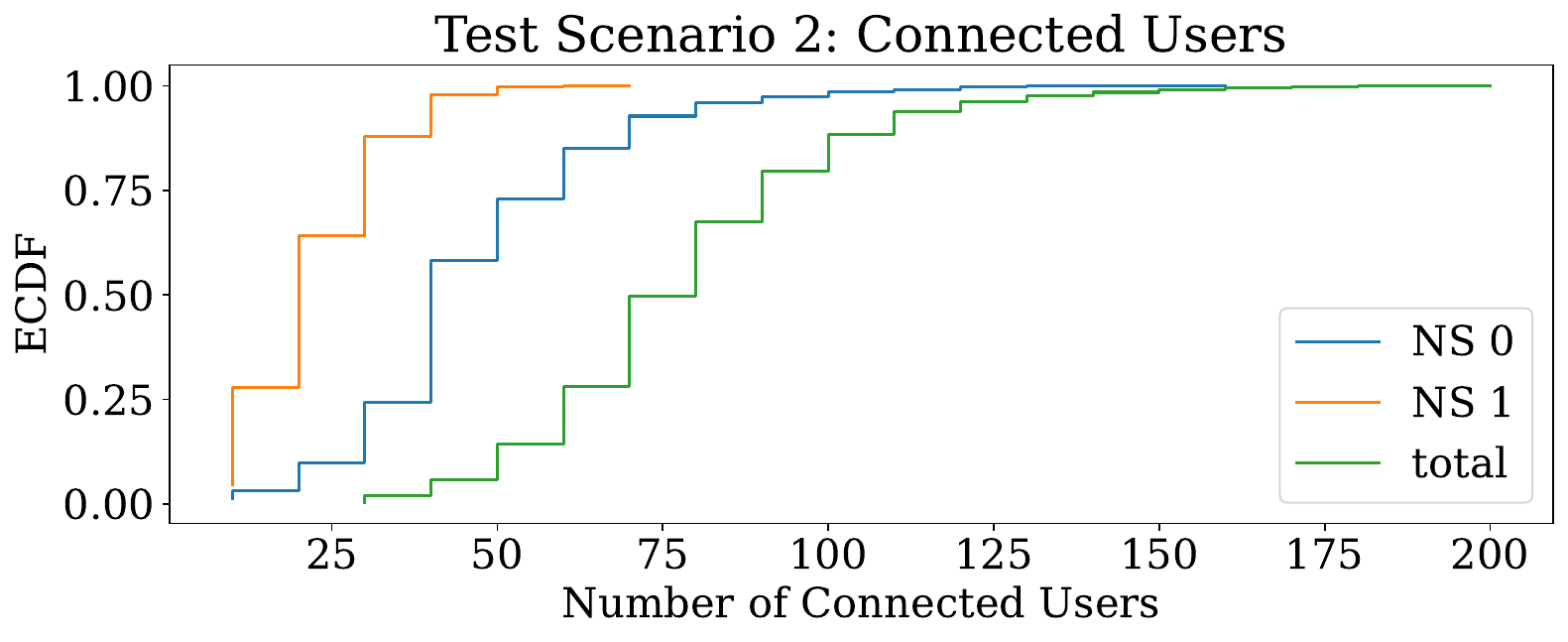}
\vspace{-0.8 cm}
\caption{The ECDF of the connected users in the regular phase.}
\label{ts2_user}
\end{figure}

\begin{figure}
\centering
\includegraphics[width=\linewidth]{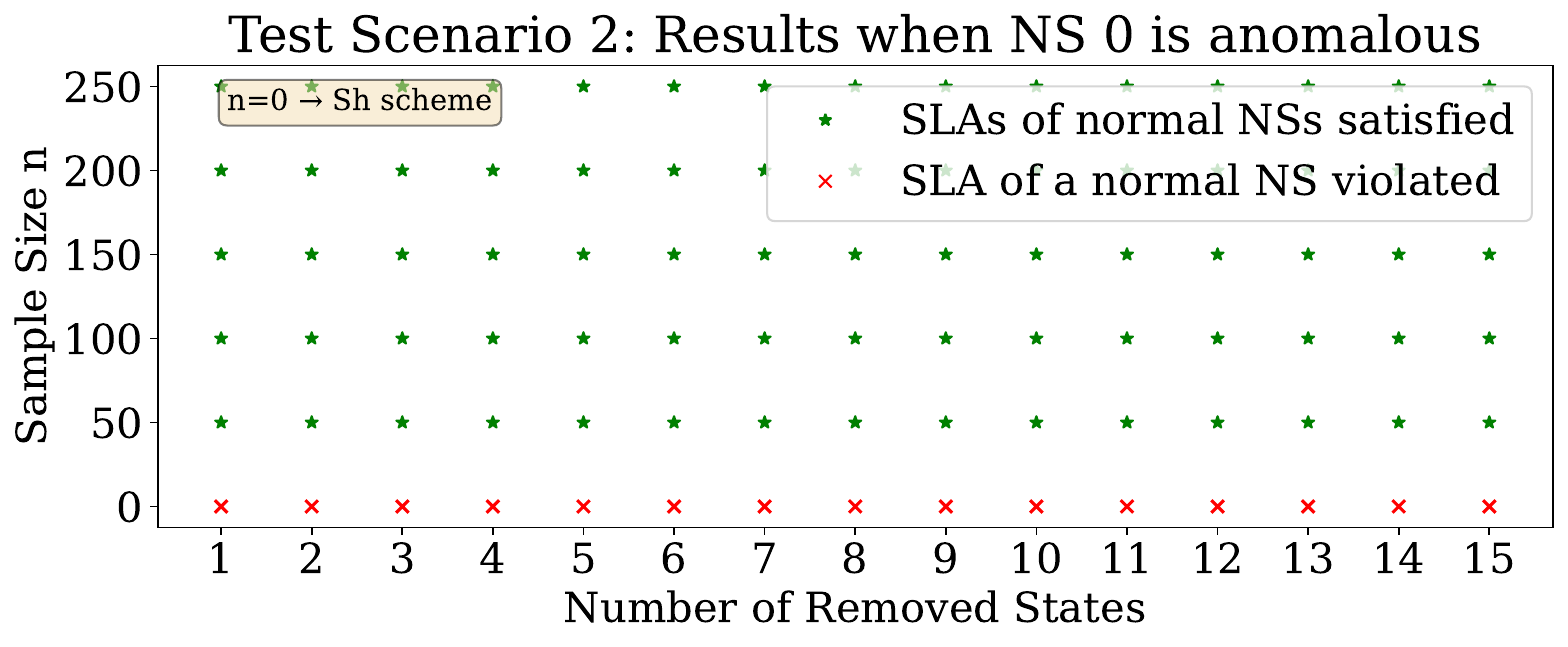}
\vspace{-0.8 cm}
\caption{The $k^{\rm{th}}$ column corresponds to case where the $k$ lowest states have been removed from the user MC of NS $0$.}
\label{ts2_aNS0}
\end{figure}

\begin{figure}
\centering
\includegraphics[width=\linewidth]{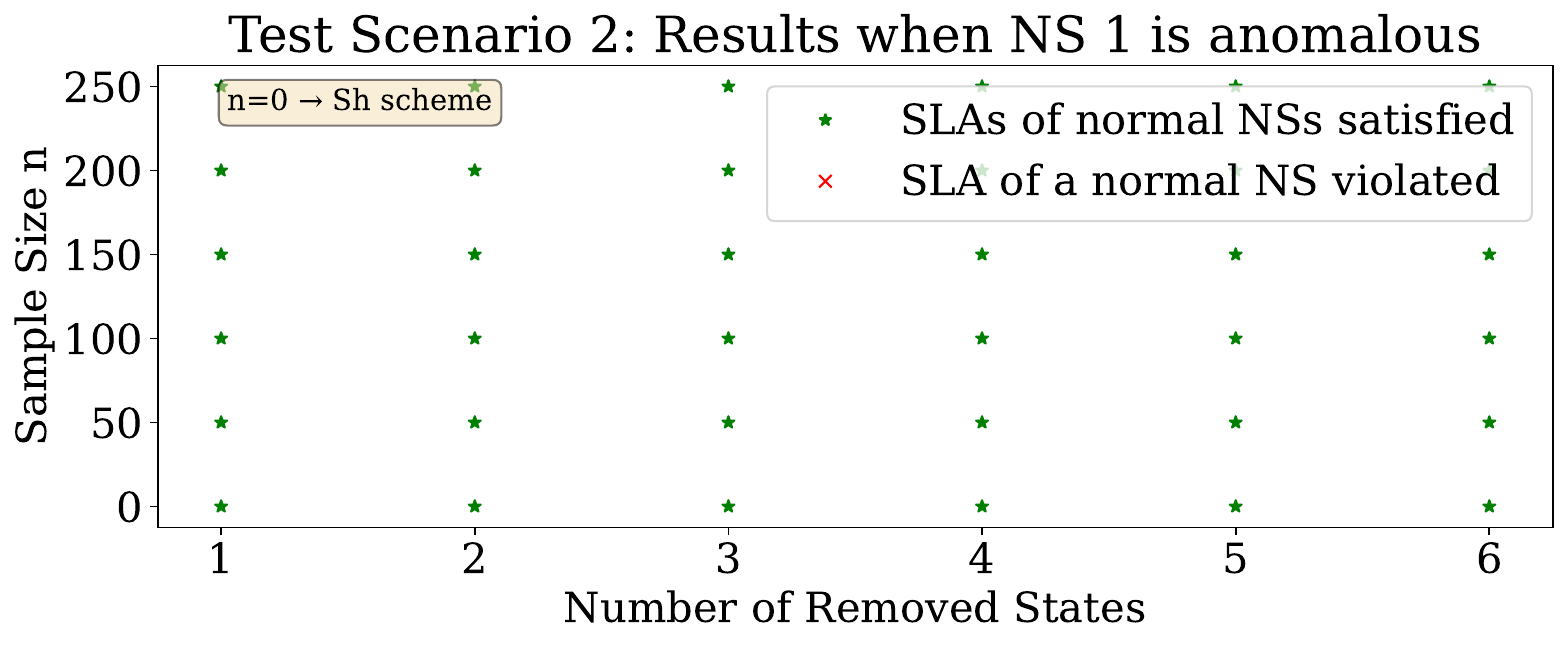}
\vspace{-0.8 cm}
\caption{The $k^{\rm{th}}$ column corresponds to case where the $k$ lowest states have been removed from the user MC of NS $1$.}
\label{ts2_aNS1}
\end{figure}

Here, we consider all possible $\beta$ for each NS to test the schemes for various anomalies. For brevity, we do not depict the granular information provided by the previous tables. In Fig. \ref{ts2_user}, we depict ECDF of the users. Table \ref{table5} shows that all schemes satisfy both SLAs when no NS misbehaves, but resource sharing schemes do so with $21\%$ less bandwidth. In Fig. \ref{ts2_aNS0}, we show the performance of the schemes as we vary the number of low states removed from the user MC of NS $0$. The results show that hypothesis testing protects the SLA of the well-behaved NS $1$. The corresponding results when NS $1$ is anomalous are shown in Fig. \ref{ts2_aNS1}. In this case, no hypothesis testing was needed, possible because the traffic in NS $1$ is lighter as shown in Fig. \ref{ts2_user}.

\begin{table}
\centering
\small
\caption{Test Scenario 2: Results when all NSs behave normally.}
\vspace{-2mm}
\begin{tabular}{lrrr}
\hline
 Scheme   &    PRBs &   $a_0$ (\%) &   $a_1$ (\%) \\
\hline
 NoSh     &    1990 &           97 &           97 \\
 Sh       &    1570 &           94 &           94 \\
 ShT50    &    1570 &           94 &           94 \\
 ShT100   &    1570 &           94 &           94 \\
 ShT150   &    1570 &           94 &           94 \\
 ShT200   &    1570 &           94 &           94 \\
 ShT250   &    1570 &           94 &           94 \\
\hline
\end{tabular}
\label{table5}
\end{table}

\subsection{Test Scenario 3}
In the third test scenario, we consider 3 NSs. The required constant bitrates are $R_0=R_2=1$ and $R_1=2$ Mbps with $P^H_0=P^H_1=P^H_2=0.9$. The data for NS $0$, NS $1$ and NS $2$ are taken from the "I-1796-raw-df-ms.parquet", "I-1815-raw-df.ms.parquet" and "II-816-raw-df.ms.parquet" files respectively. The collection period for NS $0$ is from 9:00 to 17:00 each day from June 15 to June 19, 2020. For NS $1$, we collect data from 9:00 to 17:00 each day from May 11 to May 15, 2020. Lastly, for NS $2$, the collection period is from 9:00 to 17:00 each day from April 7 to April 11, 2021.

\begin{figure}
\centering
\includegraphics[width=\linewidth]{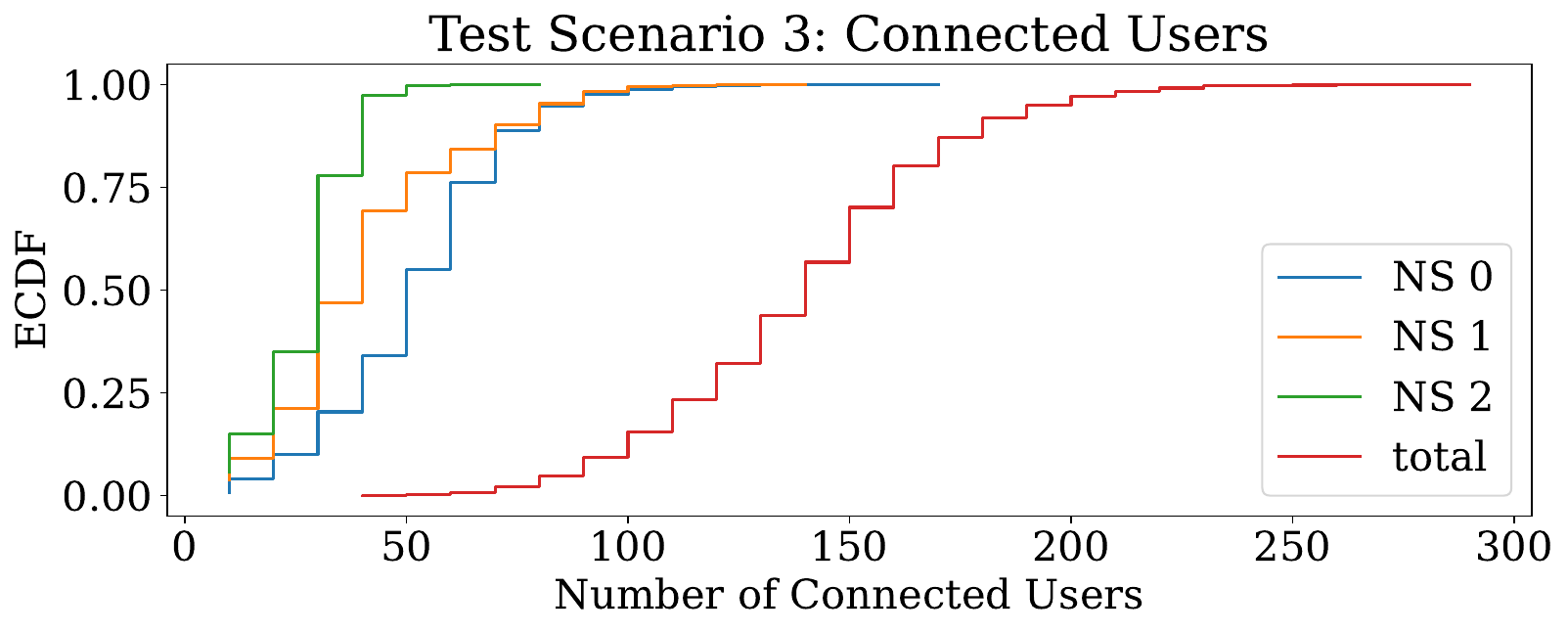}
\vspace{-0.8 cm}
\caption{The ECDF of the connected users in the regular phase.}
\label{ts3_user}
\end{figure}

\begin{figure}
\centering
\includegraphics[width=\linewidth]{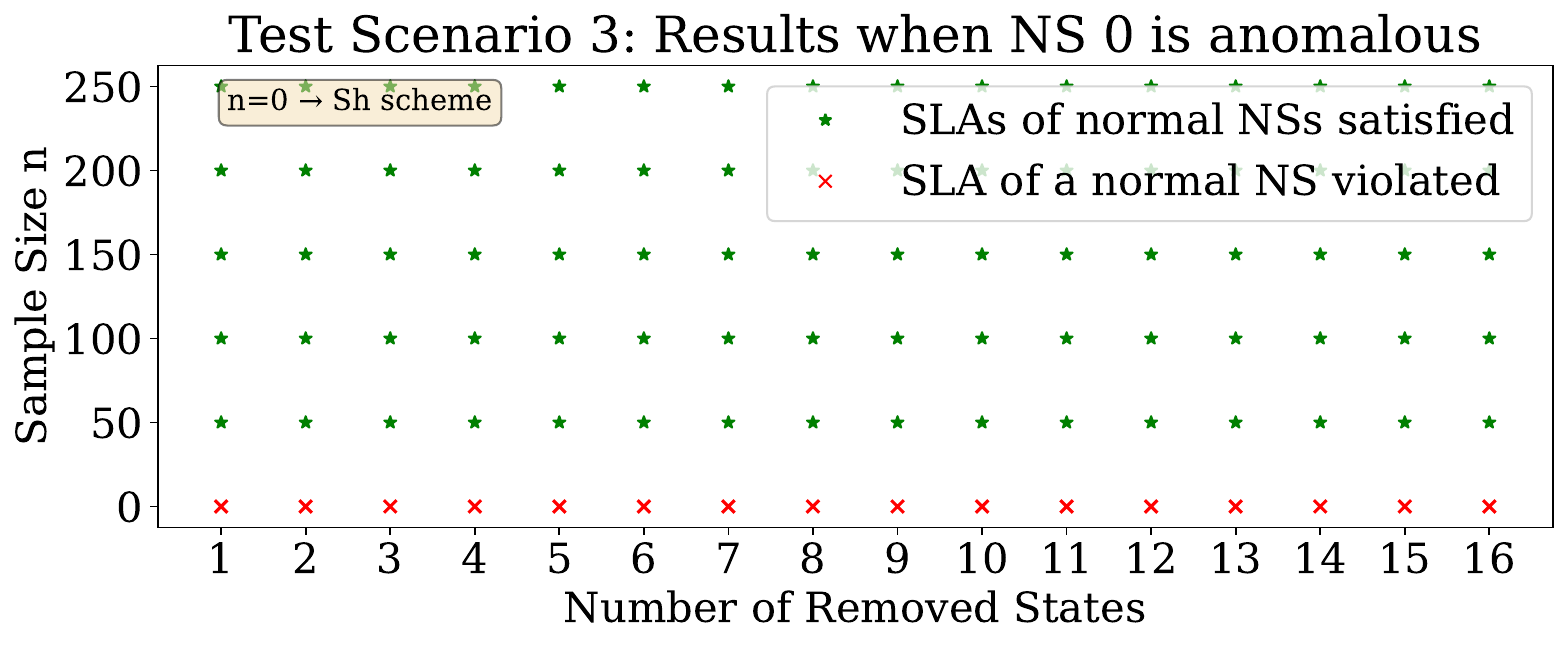}
\vspace{-0.8 cm}
\caption{The $k^{\rm{th}}$ column corresponds to case where the $k$ lowest states have been removed from the user MC of NS $0$.}
\label{ts3_aNS0}
\end{figure}

\begin{figure}
\centering
\includegraphics[width=\linewidth]{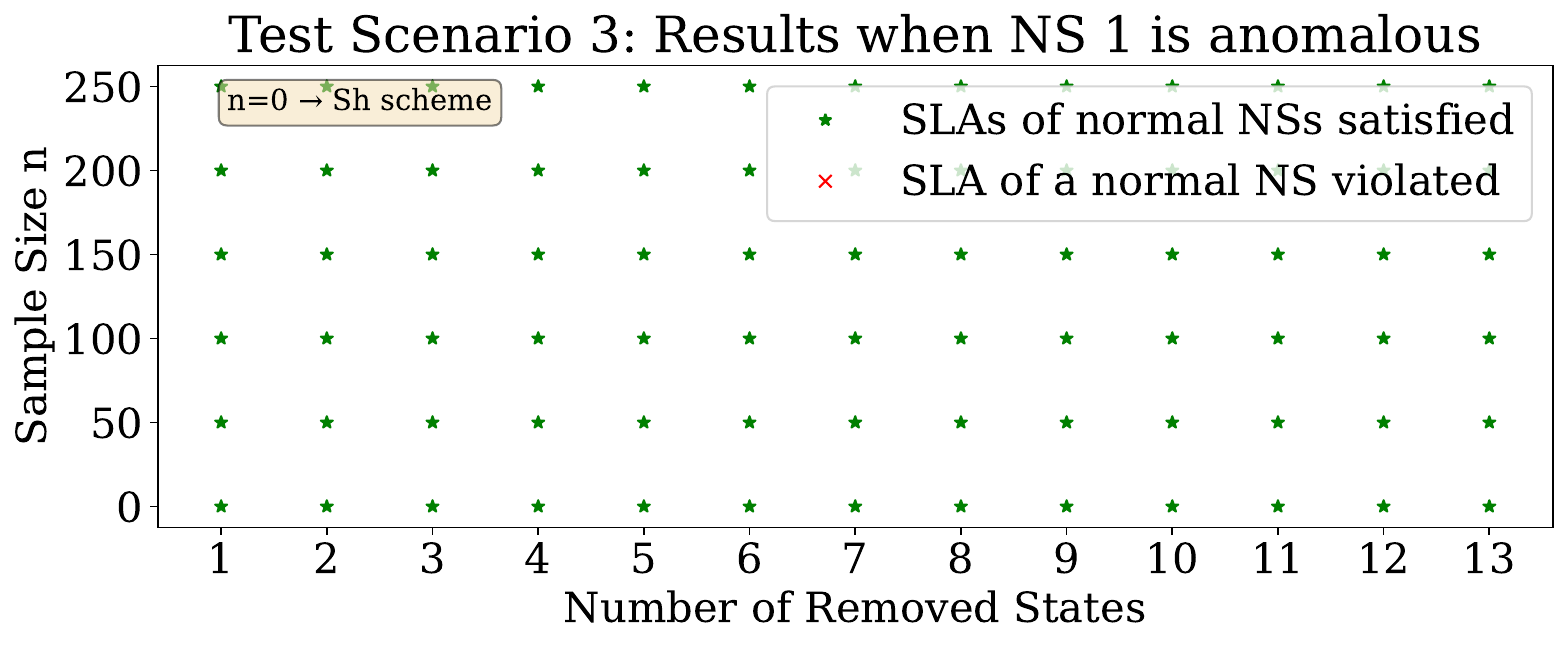}
\vspace{-0.8 cm}
\caption{The $k^{\rm{th}}$ column corresponds to case where the $k$ lowest states have been removed from the user MC of NS $1$.}
\label{ts3_aNS1}
\end{figure}

\begin{figure}
\centering
\includegraphics[width=\linewidth]{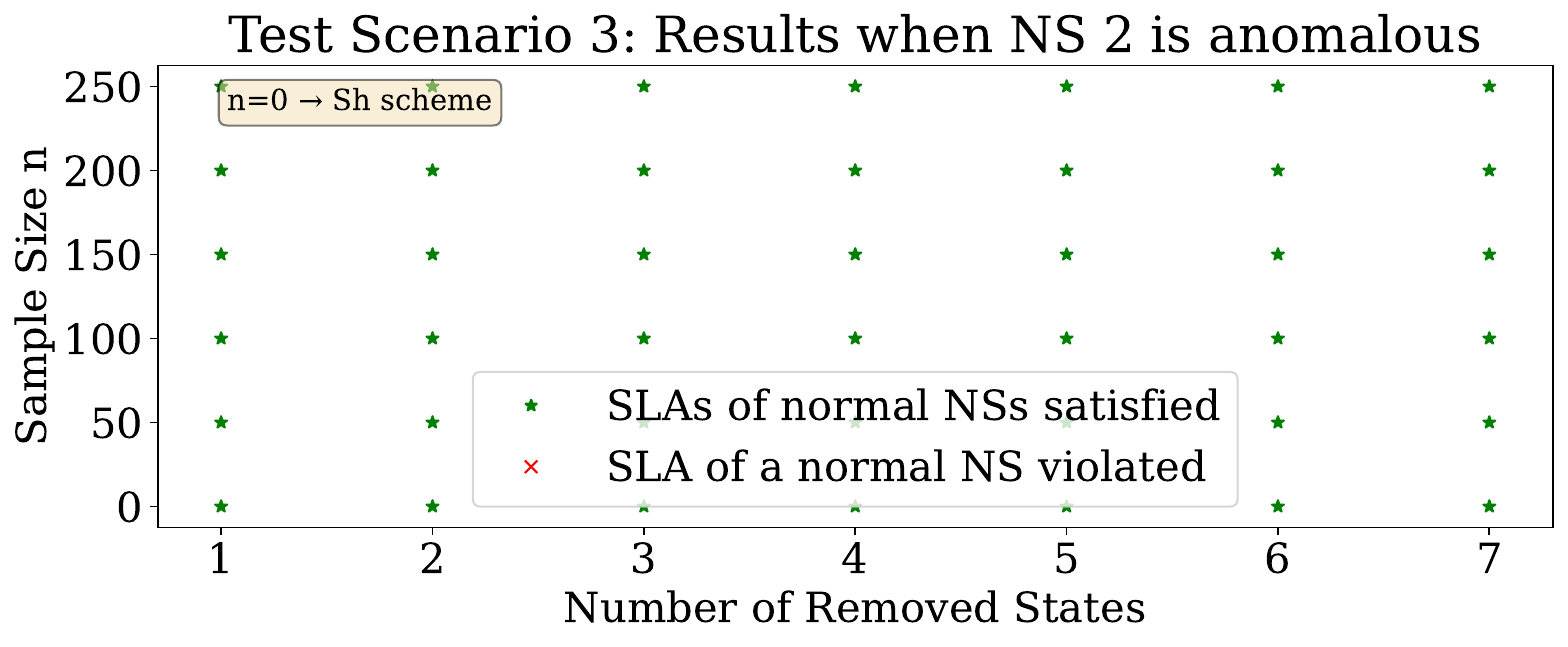}
\vspace{-0.8 cm}
\caption{The $k^{\rm{th}}$ column corresponds to case where the $k$ lowest states have been removed from the user MC of NS $1$.}
\label{ts3_aNS2}
\end{figure}

As previously, we consider all possible $\beta$ for each NS to test the schemes for various anomalies In Fig. \ref{ts3_user}, we depict the ECDF of the users for each NS. Table \ref{table6} shows that all schemes satisfy all SLAs when no NS misbehaves, but resource sharing schemes do so with $26\%$ less bandwidth. In Fig. \ref{ts3_aNS0}, we show the performance of the schemes as we vary the number of low states removed from the user MC of NS $0$. The results show that hypothesis testing protects the SLAs of the well-behaved NSs $1$ and $2$. The corresponding results when NS $1$ and NS $2$ are anomalous are shown in Fig. \ref{ts3_aNS1} and in Fig. \ref{ts3_aNS2}. In these cases, hypothesis testing is not needed.

\begin{table}
\centering
\small
\caption{Test Scenario 3: Results when all NSs behave normally.}
\vspace{-2mm}
\begin{tabular}{lcccc}
\hline
 Scheme   &    PRBs &   $a_0$ (\%) &   $a_1$ (\%) &   $a_2$ (\%) \\
\hline
 NoSh     &    2980 &           98 &           97 &           99 \\
 Sh       &    2200 &           95 &           94 &           95 \\
 ShT50    &    2200 &           95 &           94 &           95 \\
 ShT100   &    2200 &           95 &           94 &           95 \\
 ShT150   &    2200 &           95 &           94 &           95 \\
 ShT200   &    2200 &           95 &           94 &           95 \\
 ShT250   &    2200 &           95 &           94 &           95 \\
\hline
\end{tabular}
\label{table6}
\end{table}

\subsection{Time Complexity}
The two most time consuming online procedures in Algorithm \ref{regalgo} are the bandwidth allocation and hypothesis testing. The former involves the solution of a BKP which is NP-Hard. The simulations results in \cite[Fig. 3]{wiopt23} show that the BKP can be solved optimally within $5$ ms if the number of NSs is less than $100$. This can be reduced if fully polynomial time approximation schemes are used. The results were obtained using a computer with an Intel i7-10700K @3.8 GHz processor.

\begin{figure}
\centering
\includegraphics[width=\linewidth]{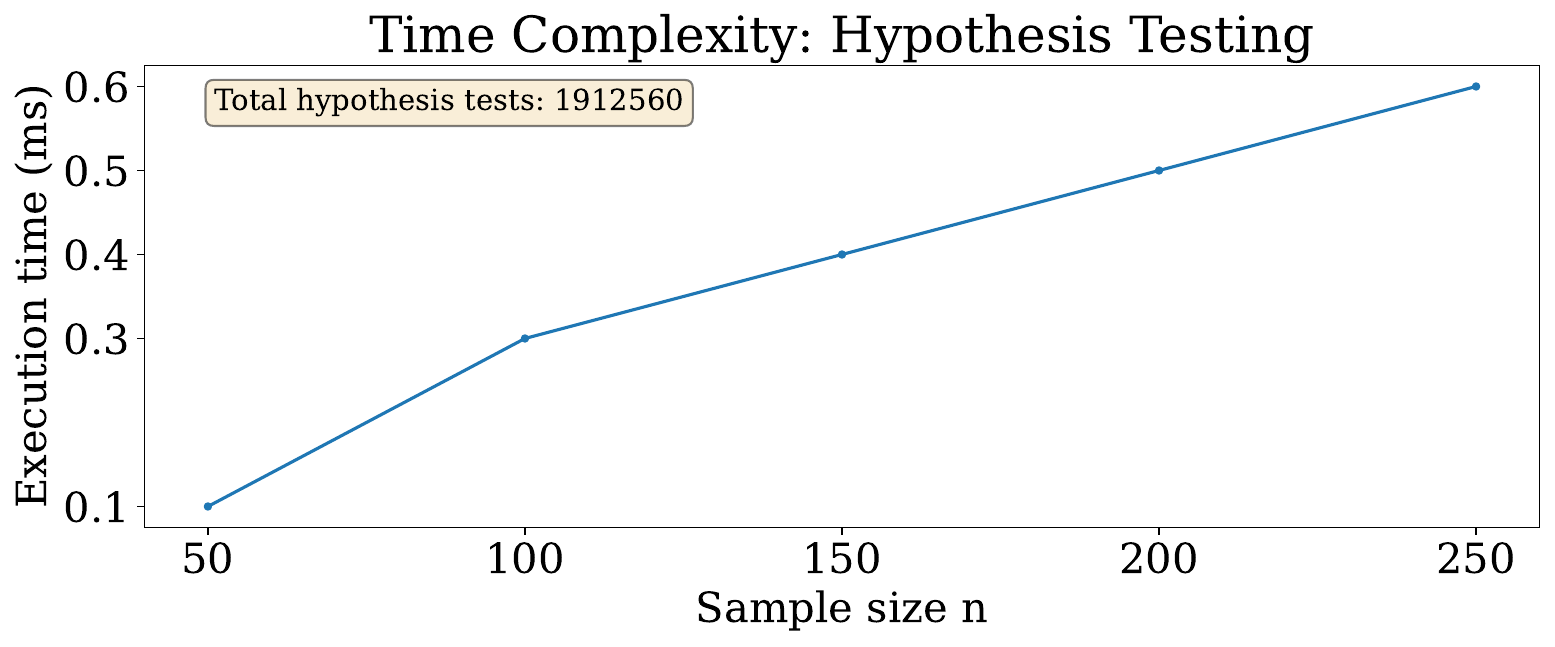}
\vspace{-0.7 cm}
\caption{The mean execution time of the hypothesis tests in test scenario 3.}
\label{execution_time}
\end{figure}

Regarding hypothesis testing, we first note that it can be performed in parallel for each NS. Therefore, it suffices to consider a single NS. Next, it is easy to see that its execution time scales linearly w.r.t. the sample size $n$. In Fig. \ref{execution_time}, we depict the mean execution time per sample size $n$ over all the tests conducted in test scenario 3. The results were obtained using a laptop with an Intel i5-7200U @2.5 GHz processor.

Overall, the total execution time of the online procedures in the regular phase are in the order of a few milliseconds. Since the slot length $D$ is in the order of tens of seconds, the proposed scheme can be performed online.

\subsection{Comments on Results}
First of all, we address the high number of PRBs reported in the previous test scenarios. This may be caused since the total number of connected users as obtained from the dataset is high and the desired QoS is to constantly provide $1$ or $2$ Mbps to each user throughout their connection time. Also, note that each NS comprises all the traffic served by a single BS.

Next, we summarize all the previous results in a single table to compare the resource efficiency and performance isolation that each scheme provides. To this end, we consider that the resource efficiency of a scheme is the percentage decrease in PRBs it achieves when compared to the no sharing scheme. Performance isolation is measured by the percentage of cases where a NS was anomalous and the SLAs of all well-behaved NSs were satisfied. Both these metrics are computed for each test scenario using the previous simulation results. Next, we combine the results by assuming that the test scenarios are equiprobable and then we report the averages in Table \ref{table7}. The table clearly shows that hypothesis testing provides both high resource efficiency and high performance isolation.

Finally, the code we developed to analyze the dataset in \cite{dataset} is available on GitHub\footnote{https://github.com/pnikolaid/robust\_resource\_sharing}. All the previous tables and figures can be reproduced using the dataset in \cite{dataset} as input to our code. 
\begin{table}
\centering
\small
\caption{Summary of Results}
\vspace{-2mm}
\begin{tabular}{lcc}
\hline
 Scheme   							&    PRB Savings (\%)	&   Performance Isolation (\%)		\\
\hline
 NoSh     					&    0 					&	100            		\\
 Sh        					&    22 				&   37         			\\
 ShT50    	&    22 				&	88           		\\
 ShT100    	&    22 				&	100           		\\ 

\hline
\end{tabular}
\label{table7}
\end{table}

\section{Conclusion}
\label{concl}
We considered the problem of satisfying the SLAs of multiple NSs. We argued that resource provisioning and dynamic resource adaptation need to be considered jointly to solve this problem. We proposed a solution approach that consists of two phases; the trial phase and the regular phase. In the trial phase, the operator estimates the required provisioned resources and obtains a model for each NS that describes its normal behavior. In the regular phase, if resource contention occurs, the operator uses the previous models to fairly decide which NSs should be rejected via hypothesis testing. Results showed that our approach is robust against traffic anomalies and satisfies the SLAs of well-behaved NSs with reduced bandwidth.

We note that there are several directions for improvement. First, a bayesian approach may be considered for hypothesis testing if priors can be estimated from past data. Alternatively, the worst case prior may be considered to formulate a minimax detection problem. Second, it may be beneficial to only check whether a NS generates more traffic than it normally does. Third, the parametric models may be used to simplify the tests. These directions may facilitate a complete performance analysis to derive the metrics in Table \ref{table7} without simulations.

\bibliography{references}

\begin{thebibliography}{10}
\providecommand{\url}[1]{#1}
\csname url@samestyle\endcsname
\providecommand{\newblock}{\relax}
\providecommand{\bibinfo}[2]{#2}
\providecommand{\BIBentrySTDinterwordspacing}{\spaceskip=0pt\relax}
\providecommand{\BIBentryALTinterwordstretchfactor}{4}
\providecommand{\BIBentryALTinterwordspacing}{\spaceskip=\fontdimen2\font plus
\BIBentryALTinterwordstretchfactor\fontdimen3\font minus
  \fontdimen4\font\relax}
\providecommand{\BIBforeignlanguage}[2]{{%
\expandafter\ifx\csname l@#1\endcsname\relax
\typeout{** WARNING: IEEEtran.bst: No hyphenation pattern has been}%
\typeout{** loaded for the language `#1'. Using the pattern for}%
\typeout{** the default language instead.}%
\else
\language=\csname l@#1\endcsname
\fi
#2}}
\providecommand{\BIBdecl}{\relax}
\BIBdecl

\bibitem{imdea}
C.~Marquez, M.~Gramaglia, M.~Fiore, A.~Banchs, and X.~Costa-Pérez, ``Resource
  sharing efficiency in network slicing,'' \emph{IEEE Trans. Netw. Service
  Manag.}, vol.~16, no.~3, pp. 909--923, 2019.

\bibitem{wiopt23}
P.~Nikolaidis, A.~Zoulkarni, and J.~Baras, ``Resource efficiency vs performance
  isolation tradeoff in network slicing,'' in \emph{2023 IEEE WiOpt}, 2023, pp.
  33--40.

\bibitem{dasilva}
C.~Sexton, N.~Marchetti, and L.~A. DaSilva, ``On provisioning slices and
  overbooking resources in service tailored networks of the future,''
  \emph{IEEE/ACM Trans. Ntw.}, vol.~28, no.~5, pp. 2106--2119, 2020.

\bibitem{zussman}
C.~Gutterman, E.~Grinshpun, S.~Sharma, and G.~Zussman, ``Ran resource usage
  prediction for a 5g slice broker,'' in \emph{ACM Mobihoc '19}, New York, NY,
  USA, 2019, p. 231–240.

\bibitem{imdea-mobihoc}
S.~Alcal\'{a}-Mar\'{\i}n, A.~Bazco-Nogueras, A.~Banchs, and M.~Fiore,
  ``kansaas: Combining deep learning and optimization for practical overbooking
  of network slices,'' in \emph{ACM MobiHoc '23}, 2023, p. 51–60.

\bibitem{survbanchs}
A.~Banchs, G.~de~Veciana, V.~Sciancalepore, and X.~Costa-Perez, ``Resource
  allocation for network slicing in mobile networks,'' \emph{IEEE Access},
  vol.~8, pp. 214\,696--214\,706, 2020.

\bibitem{isolation}
A.~J. Gonzalez~\textit{et al.}, ``The isolation concept in the 5g network
  slicing,'' in \emph{2020 EuCNC}, 2020, pp. 12--16.

\bibitem{anomaly1}
V.~Chandola, A.~Banerjee, and V.~Kumar, ``Anomaly detection: A survey,''
  \emph{ACM Comput. Surv.}, vol.~41, no.~3, jul 2009.

\bibitem{anomaly2}
------, ``Anomaly detection for discrete sequences: A survey,'' \emph{IEEE
  Trans. Knowl. Data Eng.}, vol.~24, no.~5, pp. 823--839, 2012.

\bibitem{anomaly3}
S.~Wang, J.~F. Balarezo, S.~Kandeepan, A.~Al-Hourani, K.~G. Chavez, and
  B.~Rubinstein, ``Machine learning in network anomaly detection: A survey,''
  \emph{IEEE Access}, vol.~9, pp. 152\,379--152\,396, 2021.

\bibitem{anomaly4}
M.~Gupta, J.~Gao, C.~C. Aggarwal, and J.~Han, ``Outlier detection for temporal
  data: A survey,'' \emph{IEEE Trans. Knowl. Data Eng.}, vol.~26, no.~9, pp.
  2250--2267, 2014.

\bibitem{quickchange}
V.~V. Veeravalli and T.~Banerjee, ``Chapter 6 - quickest change detection,'' in
  \emph{Academic Press Library in Signal Processing: Volume 3}.\hskip 1em plus
  0.5em minus 0.4em\relax Elsevier, 2014, vol.~3, pp. 209--255.

\bibitem{nsanomaly1}
W.~Wang, Q.~Chen, T.~Liu, X.~He, and L.~Tang, ``A distributed online learning
  approach to detect anomalies for virtualized network slicing,'' in \emph{2021
  GLOBECOM}, 2021, pp. 1--6.

\bibitem{nsanomaly2}
A.~Chawla, A.-M. Bosneag, and A.~Dalgkitsis, ``Graph-based interpretable
  anomaly detection framework for network slice management in beyond 5g
  networks,'' in \emph{2023 IEEE/IFIP NOMS}, 2023, pp. 1--6.

\bibitem{nsanomaly3}
A.~Kumar and V.~L. Thing, ``Malicious lateral movement in 5g core with network
  slicing and its detection,'' in \emph{2023 ITNAC}.\hskip 1em plus 0.5em minus
  0.4em\relax Los Alamitos, CA, USA: IEEE Computer Society, dec 2023, pp.
  110--117.

\bibitem{salman}
S.~A. Baset, L.~Wang, and C.~Tang, ``Towards an understanding of
  oversubscription in cloud,'' in \emph{USENIX Hot-ICE '12}, 2012.

\bibitem{caglar}
F.~Caglar and A.~Gokhale, ``ioverbook: Intelligent resource-overbooking to
  support soft real-time applications in the cloud,'' in \emph{2014 IEEE
  CLOUD}, 2014, pp. 538--545.

\bibitem{multitenant}
D.~Shue, M.~J. Freedman, and A.~Shaikh, ``Performance isolation and fairness
  for {Multi-Tenant} cloud storage,'' in \emph{2012 OSDI}.\hskip 1em plus 0.5em
  minus 0.4em\relax Hollywood, CA: USENIX Association, Oct. 2012, pp. 349--362.

\bibitem{3gpptables}
3GPP, ``{LTE; E-UTRA; Physical layer procedures},'' {3GPP}, TS 36.213, 02 2015,
  version 12.4.0.

\bibitem{BDE}
P.~Nikolaidis, A.~Zoulkarni, and J.~S. Baras, ``Data-driven bandwidth
  adaptation for radio access network slices,'' \emph{arXiv:2311.17347}, 2023.

\bibitem{after}
C.-Y. Hong~\textit{et al.}, ``B4 and after: Managing hierarchy, partitioning,
  and asymmetry for availability and scale in google's software-defined wan,''
  in \emph{SIGCOMM}.\hskip 1em plus 0.5em minus 0.4em\relax Budapest, Hungary:
  ACM, 2018, p. 74–87.

\bibitem{bertsekas2}
D.~P. Bertsekas, \emph{Dynamic Programming and Optimal Control, Vol. II},
  3rd~ed.\hskip 1em plus 0.5em minus 0.4em\relax Athena Scientific, 2007.

\bibitem{kay1}
S.~M. Kay, \emph{Fundamentals of statistical signal processing: estimation
  theory}.\hskip 1em plus 0.5em minus 0.4em\relax USA: Prentice-Hall, Inc.,
  1993.

\bibitem{moulos}
V.~Moulos, ``A hoeffding inequality for finite state markov chains and its
  applications to markovian bandits,'' in \emph{2020 ISIT}, 2020, pp.
  2777--2782.

\bibitem{princeton}
J.~Fan, B.~Jiang, and Q.~Sun, ``Hoeffding's inequality for general markov
  chains and its applications to statistical learning,'' \emph{J. Mach.
  Learning Research}, vol.~22, no. 139, pp. 1--35, 2021.

\bibitem{kay2}
S.~Kay, \emph{Fundamentals of statistical signal processing: Detection
  theory}.\hskip 1em plus 0.5em minus 0.4em\relax USA: Prentice-Hall, Inc.,
  1998.

\bibitem{neely}
M.~J. Neely, ``Stochastic network optimization with application to
  communication and queueing systems,'' \emph{Synthesis Lectures on
  Communication Networks}, vol.~3, no.~1, pp. 1--211, 2010.

\bibitem{ortools}
\BIBentryALTinterwordspacing
Google-OR-tools, ``The knapsack problem,'' 2023. [Online]. Available:
  \url{https://developers.google.com/optimization/pack/knapsack}
\BIBentrySTDinterwordspacing

\bibitem{dataset}
P.~F. P\'{e}rez, C.~Fiandrino, and J.~Widmer, ``Characterizing and modeling
  mobile networks user traffic at millisecond level,'' in \emph{WiNTECH
  '23}.\hskip 1em plus 0.5em minus 0.4em\relax New York, NY, USA: ACM, 2023, p.
  64–71.

\bibitem{falcon}
R.~Falkenberg and C.~Wietfeld, ``Falcon: An accurate real-time monitor for
  client-based mobile network data analytics,'' in \emph{2019 GLOBECOM}, 2019,
  pp. 1--7.

\bibitem{rnti}
G.~Attanasio, C.~Fiandrino, M.~Fiore, J.~Widmer, N.~Ludant, B.~Bloessl,
  K.~Kousias, Özgü Alay, L.~Jacquot, and R.~Stanica, ``In-depth study of rnti
  management in mobile networks: Allocation strategies and implications on data
  trace analysis,'' \emph{Computer Networks}, vol. 219, p. 109428, 2022.

\end{thebibliography}
\bibliographystyle{IEEEtran}

\end{document}